\documentclass[a4paper,11pt]{article}
\pdfoutput=1 
\usepackage{tikz}
\usetikzlibrary{matrix}

\usepackage{jcappub} 
\usepackage{bm}
\usepackage[T1]{fontenc} 

\def\x {\bm{x}}
\def\d {{d}}

\def\k {\bm{k}}
\def\x {\bm{x}}

\newcommand{\p}{_{{\text{\tiny$\|$}}}}

\newcommand{\n}{{\bf \hat{n}}}

\newcommand{\HH}{\mathcal{H}}

\newcommand{\two}{^{\text{\tiny \color{green}{({{2}})}}}}
\newcommand{\one}{^{\text{\tiny \color{red}{({{1}})}}}}

\newcommand{\<}{\langle}
\renewcommand{\>}{\rangle}
\newcommand{\dm}{{\text{{{c}}}}}

\newcommand{\HI}{\text{\tiny{HI}}}
\newcommand{\Ha}{\text{H}\alpha}

\title{\boldmath  Testing the equivalence principle on cosmological scales using  the odd multipoles of galaxy cross-power spectrum and bispectrum }

\author{Obinna Umeh,\note{Corresponding author.} }
\author{Kazuya Koyama,}
\author{Robert Crittenden.}

\affiliation{Institute of Cosmology \& Gravitation, University of Portsmouth, Portsmouth PO1 3FX, United Kingdom}

\emailAdd{obinna.umeh@port.ac.uk}
\emailAdd{kazuya.koyama@port.ac.uk}
\emailAdd{robert.crittenden@port.ac.uk }

\abstract{
One of the cornerstones of general relativity is the equivalence principle. 
However, the validity of the equivalence principle has only been established on solar system scales for standard matter fields; this result cannot be assumed to hold for the non-standard matter fields that dominate the gravitational dynamics on cosmological scales.  Here we show how the equivalence principle may be tested on cosmological scales for non-standard matter fields using the odd multipoles of the galaxy cross-power spectrum and bispectrum. This test makes use of the imprint on the galaxy cross-power spectrum and bispectrum by the parity-violating general relativistic deformations of the past-light cone, and assumes that galaxies can be treated as test particles that are made of baryons and cold dark matter.  
This assumption leads to a non-zero galaxy-baryon relative velocity 
if the equivalence principle does not hold between baryons and dark matter. We show that 
the relative velocity can be constrained to be less than {{28\%}} of the galaxy velocity using the cross-power spectrum of the HI intensity mapping/H$\alpha$ galaxy survey and the bispectrum of the H$\alpha$ galaxy survey. }

\begin{document}
\maketitle
\flushbottom


\section{Introduction}

One of the foundational pillars of general relativity is Einstein's equivalence principle.  
The equivalence principle says that all bodies fall at the same rate in a local gravitational field independently of their material make-up. Finding violations of the equivalence principle could provide clues to the nature of gravitational theories  beyond general relativity~\cite{Koyama:2015vza}. In cosmology, it could unveil the fundamental nature of dark matter and dark energy as well as the possible gravitational interaction between them~\cite{vandeBruck:2016jgg}. 

There have been various tests of the equivalence principle, starting with terrestrial studies of objects with different compositions freely falling in a vacuum ~\cite{Speake_2012,Zhou:2019bsg}. MICROSCOPE~\cite{Touboul:2017grn} tests the equivalence principle by comparing the acceleration of two masses (one made of platinum alloy and the other titanium alloy) that follow the same orbit around Earth for a long period of time\footnote{There are other tests of the equivalence principle that rely on the extra time delay that the gravitational field could cause to a propagating photon \cite{Wei:2015hwd,Giani:2020fpz} or gravitational waves~\cite{Yang:2019tzi}. These tests confirm the equivalence principle on inter-galactic scales but the level of precision of the measurement is still below the requirement for precision cosmology.}. The equivalence principle has so far been confirmed by all these tests; however, these tests  have been carried out on the solar system scale with test particles whose composition we understand\footnote{{There are constraints on an exotic coupling between ordinary and dark matter by laboratory tests of the weak equivalence when analysed as a test of the uniformity of free fall towards the centre of the Galaxy~\cite{Stubbs:1993xk}.}}. Therefore, these results cannot easily be extrapolated 
to cosmology~\cite{Aghanim:2018eyx}, where the unknown nature of cold dark matter, dark energy, gravity and backreaction is crucial to the evolution of the universe on large scales~\cite{Clarkson:2011zq,Clarkson:2011uk}.
 It is not clear whether the equivalence principle holds for these types of matter or gravity on large scales. 
 
 {Recently a method of testing the equivalence principle on cosmological scales was proposed, based on the understanding that the validity of the consistency relation between the squeezed limit of correlation functions of large scale structures is a consequence of the equivalence principle~\cite{Creminelli:2013nua,Kehagias:2013rpa}. This connection implies that any physical process that violates the equivalence principle will lead to a breakdown of the consistency relation.  Although this approach depends only on the squeezed configuration of the correlation function of galaxies, an analysis including all possible shapes of the bispectrum leads to an impressive constraint on the strength of the coupling of the galaxy velocity to the fifth force. In this paper, we describe how the number count of sources may be used to test the equivalence principle on cosmological scales by looking at the odd multipole moments of the correlation functions of the large scale structures. }
 
The number of galaxies we observe today within a redshift slice and solid angle is impacted by a number of physical effects due to the inhomogeneities along the line of sight. At leading order,  the galaxy position is displaced by the peculiar velocity of the source.  The effect of the displacement on the number count fluctuations is dominated by the redshift space distortion known as the Kaiser effect \cite{Kaiser:1987qv}. The Kaiser effect provides an avenue to test alternative theories of gravity on large scales through the measurement of the growth rate of structures~\cite{Amendola:2016saw}. The Kaiser effect induces only even multipoles of the galaxy power spectrum and bispectrum. The next-to-leading order effect on galaxy clustering is the Doppler effect, which introduces a shift in galaxy redshift whose imprint depends on the relative position of the galaxy with respect to the line of sight \cite{Bacon:2014uja,Bonvin:2015kuc,Bonvin:2016dze,DiDio:2018zmk}. The Doppler effect induces only odd multipoles of the galaxy cross-power spectrum and bispectrum~\cite{Maartens:2019yhx}. We utilise these distinct imprints of the Doppler effect on the galaxy cross-power spectrum and bispectrum to develop a consistent framework for testing the equivalence principle on cosmological scales . 

The dipole of the galaxy cross-power spectrum has been detected in the CMASS sample of the BOSS survey~\cite{Gaztanaga:2015jrs} and is seen in N-body simulations~\cite{Beutler:2020evf} by considering haloes selected based on different mass criteria. In \cite{Bonvin:2018ckp}, Bonvin and Fleury proposed to use the dipole in the galaxy cross-power spectrum to test the equivalence principle.  The approach taken in  \cite{Bonvin:2018ckp} differs substantially from the formalism we discuss here.   In particular, they proposed a parametrisation of a large class of modified Euler equation for dark matter and studied how the galaxy cross-power spectrum could constrain the parameters. 
 They also assumed that the peculiar velocities of galaxies are determined only by dark matter, and neglected the effect of baryons. 

In this paper, in addition to the dipole of the galaxy cross-power spectrum, we include the contribution from the dipole and octupole moments of the galaxy bispectrum.  We also parametrise the odd multipoles differently from Ref.~\cite{Bonvin:2018ckp}.  We do not neglect the effects of baryons,  rather we assume that galaxies can be treated as test particles that are made of cold dark matter and baryons, and baryons satisfy the standard Euler equation. This allows us to directly parametrise the relative velocity between galaxies and baryons rather than parametrising the modified Euler equation for dark matter. 
{The parametrisation we propose makes apparent the baryon-dark matter relative velocity that we are interested in.  The parametrisation is  independent of theories of gravity but we discuss  in detail a limit of this parametrisation that applies to a class of modified theories of gravity where the scale dependence of the fifth force is negligible.}
 We show that a Stage IV HI intensity mapping and the H$\alpha$ emission line galaxy survey, which overlaps in about 0.38 fraction of the sky, will be able to constrain the relative velocity of baryons and galaxies to be less than {28\%} of the galaxy velocity via the galaxy cross-power spectrum and the bispectrum of the H$\alpha$ galaxy survey.

The rest of the paper is structured as follows:  in section  \ref{sec:numbercount} we introduce the full non-perturbative expression for the number count of galaxies on arbitrary spacetime: we perturb it on an FLRW background in subsection \ref{sec:numbercountpert}, and discuss the equivalence principle for baryons in subsection \ref{sec:equivalenceprincplebaryons}. 
We adopt the standard Eulerian bias model  for the galaxy density field in section \ref{sec:models}, discussing the decomposition of the galaxy cross-power spectrum and  bispectrum in multipoles in section ~\ref{sec:multipolespowerspectrum}.  
We derive covariance matrix and describe the Fisher forecast analysis technique in section \ref{sec:Fisherforecast} and conclude in section \ref{sec:conc}. An example of how baryon-dark matter relative velocity could be sourced by an interaction in the dark sector is discussed in appendix \ref{sec:sourcesvbc}.


\noindent
{\bf{Notations}}:  
We neglect the effect of radiation and the anisotropic stress tensor, which is sufficiently accurate at $z\le 20$.
We adopt the standard normalisation for the Taylor series expansion of any quantity $X$:
$
X = \bar{X} + X\one + X\two/2
$
where $\bar{X} $ denotes the FLRW  background component, $X\one$ and $X\two$ are first and second order perturbations, respectively. We decompose each perturbed quantity at order $n$ into two parts $X^{(n)}  = X^{(n)}_{\rm{N}} + X^{(n)}_{\rm{GR}} $, where $X^{(n)}_{\rm{N}} $ denotes the Newtonian approximation of $X^{(n)}$, while $X^{(n)}_{\rm{GR}} $ denotes the general relativistic corrections. We consider the limit where only the Doppler effect dominates in $X^{(n)}_{\rm{GR}} $ and therefore  set $X^{(n)}_{\rm{GR}} = X^{(n)}_{\rm{D}} $.  
For the fiducial cosmological parameters we adopt the Planck 2018 best-fit values~\cite{Aghanim:2018eyx}: Hubble parameter, $h = 0.674$, baryon density parameter, $\Omega_b = 0.0493$, dark matter density parameter, $\Omega_{\rm{cdm}} = 0.264$, spectral index, $n_s = 0.9608$, and the amplitude of the primordial perturbation, $A_s = 2.198 \times 10^{9}$. 

\section{Galaxy number counts}\label{sec:numbercount}
The  number of galaxies seen by an observer at $o$ with a flux greater than $F$ per redshift bin and per solid angle is given by  \cite{Ellis2009,Challinor:2011bk,Alonso:2015uua}:
\begin{eqnarray}\label{eq:Numbercount}
 \frac{{\rm{dN}}^{\rm obs}(z,{\n}, F)}{{\rm{d z \,d}} \Omega_o}
 &=& \mathcal{N}_g(z,{\n}, F)\, d^2_A(z,{\n})  \left[k_{\mu} u^\mu\right]_o \bigg|\frac{\d {\lambda}}{\d{z}}\bigg|\,,
\end{eqnarray}
where  $z $ is the observed redshift of the source, $\lambda$ is the affine parameter (comoving distance) to the source, ${\n}$ is the line of sight direction to the source,  $u^\mu$ is the  4-velocity of the source galaxy, $k^{\mu}$ is the photon tangent vector,  $d_A$ is  the angular diameter distance to the source and $\mathcal{N}_g$ is the flux-limited proper number density of galaxy
\begin{eqnarray}\label{eq:fluxlimit}
\mathcal{N}_g(z,{\n}, F) = \int_{ \ln L(F)}^{\infty} \d  \ln L\; n_g(z,{\n}, ]\ln L)\,.
\end{eqnarray}
Here $n_g$ is the proper number density of sources. The luminosity of the source is related to its flux by 
$ L=4\pi {F} d_L^2= 4\pi {F}(1+z)^4 d_A^2$.

The expansion of equation \eqref{eq:Numbercount} up to second order in perturbation theory  has been done by several authors \cite{Bertacca:2014hwa,DiDio:2015bua,Yoo:2014sfa}. However, they all assumed that the motion of galaxies traces that of dark matter and that dark matter obeys the  equivalence principle.  This is the key assumption we relax here. 

We consider metric perturbations in Poisson gauge on a background FLRW spacetime:
\begin{eqnarray}\label{eq:Poisson-gauge}
\d s^2 &=&a^2\left( -\left(1+2\Phi\right)\d\eta^2+ \left(1-2\Psi\right)\delta_{ij}\d{x^i}\d x^j\right) \,,
\end{eqnarray}
where $\Phi$ and $\Psi$ are the metric perturbations, ${``a"}$ is the scale factor of the background  FLRW spacetime, $\eta$ is the conformal time.  We neglect the vector and tensor perturbation at first and second order, because they are sub-dominant in this gauge~\cite{Jolicoeur:2018blf}.
For a given fluid component $I$, the perturbation of the temporal and spatial components of its 4-velocity is given by
\begin{eqnarray}
u^0_{\tiny{I}}&=&1 - \Phi\one +  \frac{1}{2}\left[3 [\Phi\one]^2 - \Phi\two + {\partial}_{i}v_{I}\one{\partial}^{i}v_{I}\one\right]  \,,\\
u^i_{\tiny{I}}&=&{\partial}^{i}v\one_{\tiny{I}} + \frac{1}{2}  {\partial}^{i}v\two_{\tiny{I}}\,,
\end{eqnarray}
where  $v_{I}$ is the $I$-th peculiar velocity potential..  We shall see later how the peculiar velocity potential of each of the fluid components is related to $\Phi$ via the generalised Euler equation. 

\subsection{Gravity theory independent number count in the weak-field limit}\label{sec:numbercountpert}
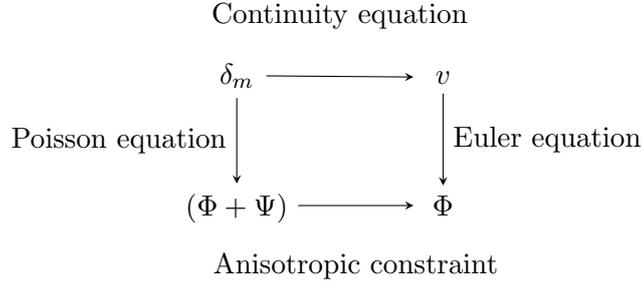
\begin{figure}[h]
\begin{center}
\begin{tikzpicture}
  \matrix (m) [matrix of math nodes,row sep=3em,column sep=4em,minimum width=2em]
  {
     \delta_m & v \\
   (\Phi +\Psi) & \Phi \\};
  \path[-stealth]
    (m-1-1) edge node [left] {${\rm{Poisson~equation}}$} (m-2-1)
            edge node [above =0.5cm] {${\rm{Continuity~equation}}$} (m-1-2)
    (m-2-1.east|-m-2-2) edge node [below =0.5cm] {${\rm{Anisotropic~constraint}}$}
            node [above] {} (m-2-2)
    (m-1-2) edge node [right] {${\rm{Euler~equation}}$} (m-2-2);
\end{tikzpicture}
\end{center}
\caption{Scalar perturbations in Poisson-gauge have four independent degrees of freedom: two from the matter sector, $\delta_m$(matter density), $v$ (peculiar velocity) and two from the metric sector, $\Phi$ (gravitational potential), and $\Psi$ (curvature perturbation). Large scale structure survey measures $\delta_m$ via clustering of biased tracers, $v$ via the redshift space distribution, $\Phi$ via the Doppler effect and $\Phi +\Psi$ via weak gravitational lensing. }
\label{fig:relationship}
\end{figure}

In the weak-field limit, we neglect integrated terms such as weak gravitational  lensing, the integrated Sachs-Wolfe effect, the time-delay effect, etc. The contributions from such terms are expected to be sub-dominant for thin redshift bins, although they could be important when cross-correlations between widely separated redshift bins are considered \cite{Montanari:2015rga,Jelic-Cizmek:2020pkh}.
We shall introduce the basic notations here and refer the reader to \cite{Umeh:2014ana,Bertacca:2014hwa} for details on the derivation. Expanding equation \eqref{eq:fluxlimit} in perturbation theory leads to 
 \begin{eqnarray}
\mathcal{N}_g(z,{\n},\bar{L}) &=&\bar{\mathcal{N}}_g(z, \bar L) \bigg[1 + \delta_g
   +b_e \Delta_z
 +{\mathcal{Q}}  \Delta_{d_L}+\frac{ \partial \delta_{g} }{\partial\ln \bar L}\Delta_{d_L}\bigg]\,,
   \end{eqnarray}
where we have introduced the following short-hand notations for simplicity:
the perturbation in the galaxy number density  $\delta_g\equiv\delta_g(z,{\n},\bar L)  $,  the magnification bias parameter ${\mathcal{Q}} \equiv{\mathcal{Q}}(z,\bar L)$, which is related to the slope of the luminosity function $s$ according to  $\mathcal{Q} =5 s/2$, the evolution bias parameter $b_e \equiv b_e(z,\bar L)$, the redshift  perturbation $\Delta_z \equiv\Delta_z(z,{\n})$ and the perturbation of the luminosity distance $ \Delta_{d_L}\equiv \Delta_{d_L}(z,{\n})$.
Putting all these in equation \eqref{eq:Numbercount}  and expanding everything up to second order leads to 
 \begin{eqnarray}
  \frac{{\rm{d N}}^{\rm obs}(z,\n,F)}{{\rm{d z}}\, {\rm{d}} \Omega_o} & =& \frac{{\rm{d \bar{N}}}(z,F)}{{\rm{d z}}\, {\rm{d \Omega}}_o}\bigg[1  +   \Delta\one_{\rm{N}^{\rm obs}} (z,{\n},F) + \frac{1}{2}  \Delta_{\rm{N}^{\rm{obs}}}\two(z,{\n},F) \bigg]  \,,
 \label{eq:NumberDA}
  \end{eqnarray}
where  ${{\rm{d\bar{N}}}(z,F)}/{{\rm{d z}}\, {\rm{d \Omega}}_o}$ is the mean number count per redshift bin per solid angle  and $\Delta_{\rm{N}^{\rm obs}}$ is fluctuation. We can write $ \Delta_{\rm{N}^{\rm obs}}  = \Delta_{\rm{N} } +   \Delta_{\rm{D}}$, where at linear order 
 \begin{eqnarray} \label{eq:linearordergalaxydensityKaiser}
\Delta\one_{\rm{N} } &=& \delta_{g}\one- \frac{1}{\HH}  \partial_{\|}^2 {v_g}\one\,,
   \\
   \Delta\one_{\rm{D} } &=& \partial_{\p} {v_g}\one + \frac{1}{\HH} \left( \partial\p {{v_g}\one}' + \partial\p \Phi\one\right)
  + \left[ b_e   - 2 \mathcal{Q}  - \frac{2\left(1- \mathcal{Q}\right)}{\chi \HH} - \frac{\HH'}{\HH^2} \right] \partial_{\|} {v_g}\one \,, 
   \label{eq:linearordergalaxydensityDoppler}
\end{eqnarray}
where $\partial_{\|}^2 {v_g}\one =n^in^j\partial_i\partial_j {v_g}\one$.
The equation \eqref{eq:linearordergalaxydensityKaiser} is the well-known Kaiser limit \cite{Kaiser:1987qv} of the number count fluctuations, which constitutes what we call the Newtonian approximation $\Delta\one_{\rm{N} } $.  Equation \eqref{eq:linearordergalaxydensityDoppler}  contains the leading order contribution to the large-scale general relativistic effects;  $ \Delta\one_{\rm{D}}$ is dominated by the Doppler effects \cite{Clarkson:2018dwn}.
Equation \eqref{eq:linearordergalaxydensityDoppler}  is independent of any theory of gravity. At second order, we find
\begin{eqnarray}
  \Delta\two_{\rm{N}} &=& \delta_{g}\two- \frac{1}{\HH}  \partial_{\|}^2 {v_g}\two  - \frac{2}{\mathcal{H}}\bigg[\delta_{g}\one\partial_{\parallel}^{2}{v_g}\one + \partial_{\parallel}{v_g}\one\partial_{\parallel}\delta_{g}\one\bigg] 
  \label{eq:Newtoniannumbercount2}
 + \frac{2}{\mathcal{H}^{2}}\bigg[\left(\partial_{\parallel}^{2}{v_g}\one\right)^{2} 
 \nonumber \\  && 
 + \partial_{\parallel}{v_g}\one\partial_{\parallel}^{3}v_{g}\one\bigg] \,,
 \\
  \Delta\two_{\rm{D}} &=& 
\left[1+ b_e   - 2 \mathcal{Q}  - \frac{2\left(1- \mathcal{Q}\right)}{\chi \HH} - \frac{\HH'}{\HH^2} \right] \partial_{\|} {v_g}\two + \frac{1}{\HH} \left( \partial\p {{v_g}\two}' + \partial\p \Phi\two\right) \nonumber
  \\ \nonumber && 
+\frac{2 }{\HH}\left[\delta_g\one-\frac{2}{\HH}\partial\p^2 {v_g}\one \right] \left[ \partial\p {{v_g}\one}' + \partial\p \Phi\one \right]  +\frac{2}{\HH} \Phi\one\left[\partial\p \delta_g\one - \frac{1}{\HH}\partial^3\p {v_g}\one\right]
\\ \nonumber &&
 + 4\partial\p {v_g}\one  \left( 1 - \frac{1}{\chi \HH}\right) \frac{\partial \delta_g\one}{\partial \ln L}  +   2\partial\p {v_g}\one \delta_g\one \left[1+ b_e   - 2 \mathcal{Q}  - \frac{2\left(1- \mathcal{Q}\right)}{\chi \HH} - \frac{\HH'}{\HH^2} \right] 
\\ \nonumber &&
  +\frac{2}{\HH}\partial\p {{v_g}\one}\left[{\delta_g\one}' - \frac{{2}}{\HH}\partial^2\p {{v_g}\one}' -\frac{1}{\HH}\partial\p^2 {\Phi\one}\right]
  - \frac{2}{\HH} \nabla_{\bot i} {v_g}\one \nabla_{\bot}^i \partial\p {v_g}\one  
\\  &&  
+ \frac{2}{\HH} \partial\p {v_g}\one\partial\p^2 {v_g}\one\left[-2 - 2b_e   + 4 \mathcal{Q}  + \frac{4\left(1- \mathcal{Q}\right)}{\chi \HH} + 3\frac{\HH'}{\HH^2} \right] \,.
\label{eq:Secondordergalaxydensity0}
\end{eqnarray}
{Furthermore, equation \eqref{eq:linearordergalaxydensityDoppler} contains the  gradient of the gravitational potential, while equation \eqref{eq:Secondordergalaxydensity0} contains both the gravitational potential and the gradient of the gravitational potential. In order to express them in terms of the Doppler peculiar velocity, the Euler equation is required. See Figure \ref{fig:relationship} for the relationships between scalar perturbation variables.}

 {The classification of the second order terms into Newtonian and Doppler terms is a little more complicated. However, we found that the most consistent way to classify all the terms is to think of the density term and the velocity term as a  function of the gravitational potential through the Poisson equation (in sub-horizon limit) and the baryon Euler equation, respectively. Within this scheme, the quadratic second-order Newtonian terms will be proportional to four spatial derivatives of two gravitational potentials. How the derivatives act on the gravitational potentials does not matter. Similarly, the quadratic second-order Doppler terms in equation \eqref{eq:Secondordergalaxydensity0} contain only terms with three spatial derivatives of the two gravitational potentials. Again, how the derivatives act on the gravitational potentials does not matter, what is important is the number of spatial derivatives.}

Galaxies are made of baryons and dark matter.  In the standard case, both baryons and dark matter are assumed to follow the same geodesic equation
\begin{equation}\label{eq:matterEuler}
{\partial^i {v_{m}}\one}' + \HH \partial^i{v_{m}}\one+{\partial^i}\Phi\one=0\,.
\end{equation} 
If the galaxy velocity exactly coincides with the matter velocity $v_{g} = v_{m}$,  we can use equation \eqref{eq:matterEuler} and its second order equivalent to relate  $\Phi$ to ${v_{g}}$, and the result is the well-known expression for the general relativistic number count fluctuation given in \cite{Challinor:2011bk,Bonvin:2011bg,Yoo:2010ni} at linear order and  in \cite{Bertacca:2014hwa,DiDio:2015bua,Yoo:2014sfa,DiDio:2020jvo} at second order.

\subsection{Beyond the standard case: Equivalence principle for baryons only}\label{sec:equivalenceprincplebaryons}

 We assume that only the baryon motion is geodesic, hence they satisfy the standard  Euler equation. At the linear order, the Euler equation for baryons is given by 
\begin{eqnarray}\label{eq:Eulerianforbaryons1}
{\partial^i {v_{b}}\one}' + \HH \partial^i{v_{b}}\one+{\partial^i}\Phi\one=0 \,,
\end{eqnarray}
where $v_b\one$ is the linear  baryon peculiar velocity potential. 
Using equation \eqref{eq:Eulerianforbaryons1} in equation \eqref{eq:linearordergalaxydensityDoppler} to relate the gravitational potential to the baryon velocity. we find 
\begin{eqnarray}\label{eq:linearordergalaxydensityDoppler2}
  \Delta\one_{\rm{D}} &=&
   \partial_{\p} {v_g}\one - \partial_{\p} {v_b}\one 
      \\ \nonumber &&
      + \frac{1}{\HH} \left( \partial\p {{v_g}\one}' - \partial\p {{v_b}\one}'\right)
  + \left[ b_e   - 2 \mathcal{Q}  - \frac{2\left(1- \mathcal{Q}\right)}{\chi \HH} - \frac{\HH'}{\HH^2} \right] \partial_{\|} {v_g}\one \,.
\end{eqnarray}
Similarly, the Euler equation for baryons at second order in the weak field limit is given by
\begin{eqnarray}
\partial_i {v_{b}\two}' & +& \HH\partial_i {v_{b}\two}+   \partial_i\Phi\two  + 2 \partial_i \partial_j v_{b}\one \partial^jv_{b}\one   
 +\mathcal{O}(\Phi\one\partial\Phi)= 0\,.
   \label{eq:Eulerianforbaryons2}
\end{eqnarray}
Using equations \eqref{eq:Eulerianforbaryons1} and \eqref{eq:Eulerianforbaryons2}, we can relate $\Phi\one$ to $v_{b}\one$ and $\Phi\two$ to $v_{b}\two$ in equation \eqref{eq:Secondordergalaxydensity0}.  After straight-forward  but lengthy algebra we find
\begin{eqnarray}\nonumber
  \Delta\two_{\rm{D}} &=& 
\left[ b_e   - 2 \mathcal{Q}  - \frac{2\left(1- \mathcal{Q}\right)}{\chi \HH} - \frac{\HH'}{\HH^2} \right] \partial_{\|} {v_g}\two +
( \partial_{\|} {v_g}\two-\partial_i {v_{b}\two}) + \frac{1}{\HH} \left( \partial\p {{v_g}\two}' -\partial_i {v_{b}\two}' \right)
  \\ \nonumber && 
 -\frac{2}{\HH}  \partial_{\p} \partial_j v_{b}\one \partial^jv_{b}\one  
+\frac{2 }{\HH}\left[\delta_g\one-\frac{2}{\HH}\partial\p^2 {v_g}\one \right] \left[ \partial\p {{v_g}\one}' -{\partial_{\p} v_{b}\one}' - \HH  \partial_{\p}v_{b}\one \right] 
  \\ \nonumber && 
 -\frac{2}{\HH} \left[{v_{b}\one}' +\HH  v_{b}\one\right]\left[\partial\p \delta_g\one - \frac{1}{\HH}\partial^3\p {v_g}\one\right]
\\ \nonumber &&
 + 4\partial\p {v_g}\one  \left( 1 - \frac{1}{\chi \HH}\right) \frac{\partial \delta_g\one}{\partial \ln L}  +   2\partial\p {v_g}\one \delta_g\one \left[1+ b_e   - 2 \mathcal{Q}  - \frac{2\left(1- \mathcal{Q}\right)}{\chi \HH} - \frac{\HH'}{\HH^2} \right] 
\\ \nonumber &&
  +\frac{2}{\HH}\partial\p {{v_g}\one}\left[{\delta_g\one}' - \frac{{2}}{\HH}\partial^2\p {{v_g}\one}' +\frac{1}{\HH}{\partial^2_{\p} v_{b}\one}' +   \partial^2_{\p}v_{b}\one\right]
  - \frac{2}{\HH} \nabla_{\bot i} {v_g}\one \nabla_{\bot}^i \partial\p {v_g}\one  
\\  &&  
+ \frac{2}{\HH} \partial\p {v_g}\one\partial\p^2 {v_g}\one\left[-2 - 2b_e   + 4 \mathcal{Q}  + \frac{4\left(1- \mathcal{Q}\right)}{\chi \HH} + 3\frac{\HH'}{\HH^2} \right] \,.
\label{eq:Secondordergalaxydensity2}
\end{eqnarray}
We parametrise the relative velocity between galaxy peculiar velocity $v_g$ and baryon peculiar velocity $v_b$ as 
\begin{equation}\label{eq:veldiffparam}
v_g\one - v_b\one  = v\one_{gb} \equiv \Upsilon_1  v_g\one, \quad 
v_g\two - v_b\two  =v\two_{gb}\equiv \Upsilon_{2}  v_g\two\,.
\end{equation}
{We  have introduced two spacetime dependent parameters $\Upsilon_1=\Upsilon_1({\eta,{\x}}) $ and $ \Upsilon_2=\Upsilon_2({\eta,{\x}}) $ which modulate the  linear order  galaxy-baryon relative velocity, $v\one_{gb}$, and the second order  galaxy-baryon  relative velocity, $v\two_{gb}$, respectively. } This might appear to be introducing too many parameters to describe the same physical effect; however, this is not the case especially for modified gravity theories where the second order dark matter density field and peculiar velocity satisfy  non-linear second order differential equations, thus its evolution could be significantly different from the linear order expression as shown in Appendix A.  
 We  parametrise the derivative with respect to conformal time as 
\begin{equation}
{{v_g}\one}' - {{v_b}\one}' ={v_{gb}\one}' = \beta_1  {v_g\one}', \quad 
{{v_g}\two}' - {{v_b}\two}' ={v_{gb}\two}' = \beta_2  {v_g\two}'\,.
\end{equation}
{For simplicity, we assume that $ \Upsilon_1$ and $\Upsilon_{2}$ are a smooth and well-behaved functions of the conformal time only.  In this limit, $\{\beta_1 ,\beta_2\}$ may be expressed in terms of $\{\Upsilon_1,\Upsilon_{2}\}$;}
\begin{eqnarray}\label{eq:fixbeta12}
\beta_1 &=& \frac{\Upsilon'_{1}}{\bar{X}_{g1} } + \Upsilon_{1}\,,
\qquad 
\beta_2=\frac{\Upsilon'_{2}}{ \bar{X}_{g2} } + \Upsilon_{2}\,,
\end{eqnarray}
where $\bar{X}_{g 1} \equiv{  {v_g}\one}'/{{v_g}\one}$ and $\bar{X}_{g 2} \equiv{ {v_g}\two}'/{{v_g}\two}$. 
With these, we can express the galaxy number count in terms of $\delta_g$ and $v_g$ only. {The only assumptions we have made so far is that baryons obey the standard Euler equation (the equivalence principle) and that the galaxy-baryon relative velocity is a function of the observed galaxy velocity with a time-dependent amplitude only. 
We note that scale dependence may arise from the mass of the scalar field that mediates the fifth force. Scale-dependent growth will be better constrained by the even multipole moments of the power spectrum and bispectrum. In the rest of the paper, we assume that scale dependence can be ignored on scales relevant to our analysis.}  

\section{Model independent constraint on the relative velocity}\label{sec:models}
We assume that the galaxy density is well described in terms of the total matter density $\delta_m$ and the scalar invariant of the  tidal field tensor $\mathcal{S}^2$ according to  
\begin{eqnarray}
\delta_{\rm{g}}&=& b_1\delta_{m}\one 
+ \frac{1}{2}\bigg[
b_1\delta_{m}\two + 
b_2(\delta_{\rm{m}}\one)^2 + b_{s^2}\mathcal{S}^2 \bigg]
\,,\label{eq:Eulerianbiasmodel}
\end{eqnarray}
where $b_1$, $b_2$ and $ b_{\mathcal{S}^2}$ are linear, non-linear and tidal bias, respectively. 
In principle, the existence of a non-vanishing $v_{gb}$ could contribute extra terms to the galaxy bias formula given in equation  \eqref{eq:Eulerianbiasmodel}.  (For the case of $v_{gb}$ sourced during the photon-Baryon decoupling, preliminary studies have indicated that the effect of such terms is negligible for a Stage IV large scale structure survey \cite{Tseliakhovich:2010bj,Barreira:2019qdl}.) 
For our purposes, we can ignore such terms because the galaxy density will be directly constrained by the even multipoles of the N-point correlation function \cite{Schmidt:2016coo,Barreira:2019qdl}. Similarly, we neglect the general relativistic corrections to equation \eqref{eq:Eulerianbiasmodel} since they become important near Hubble horizon scales \cite{Umeh:2019qyd}, whose consideration is beyond the scope of this work. 

The matter density at first order is given by $\delta\one_{m}({\k},\eta) = D_{m}(\eta) \delta\one_{O}({\k})\,,$  where  $D_{m}$ is the growth of the matter density field ({for simplicity, we assume that it is a function of the conformal time only}.), $\delta_{O}({\k}) $ is related to the initial density field via the transfer function $T(k)$ :  $\delta_{O}({\k}) = \delta_{\rm{ini}}({\k}) T(k)$.  At second order the matter density and tidal tensor invariant are given by 
\begin{eqnarray}
\delta_{m}\two({\k},\eta)&=& \int \frac{d^3k_1}{(2\pi)^3}\frac{d^3k_2}{(2\pi)^3}F_{2} ({\k}_{1}, {\k}_{2})\delta_{m}({\k}_1, \eta)\delta_{m}({\k}_2, \eta)(2\pi)^3 \delta^{(3)}\left({\k} -{\k}_1 -{\k}_2\right)
\,, \\
\mathcal{S}^2({\k},\eta) &=&  \int \frac{d^3k_1}{(2\pi)^3}\frac{d^3k_2}{(2\pi)^3}S_{2} ({\k}_{1}, {\k}_{2})\delta_{m}({\k}_1, \eta)\delta_{m}({\k}_2, \eta)(2\pi)^3 \delta^{(3)}\left({\k} -{\k}_1 -{\k}_2\right)\,,
\end{eqnarray}
where $F_2$ and $S_2$ are the  Fourier space kernels of the matter density field and the scalar invariant of the tidal field tensor, respectively.
Similarly, we model the galaxy  velocity  in terms of the matter density field as follows
\begin{eqnarray}\label{eq:galaxyoverfnl1}
v_{g}\one({\k},\eta) &=& \frac{\HH}{k^2} f_{g}\delta_{m}\one({\k},\eta)\,,
\\ 
\label{eq:galaxyoverfnl2}
v_g\two({\k},\eta)&=&\frac{\HH f_{g}}{k^2} \int \frac{d^3k_1}{(2\pi)^3}\frac{d^3k_2}{(2\pi)^3}{G_2} ({\k}_{1}, {\k}_{2})
\delta_{m}({\k}_1, \eta)\delta_{m}({\k}_2, \eta)  \delta^{(3)}\left({\k} -{\k}_1 -{\k}_2\right)
\,,
\end{eqnarray}
where $f_{g}$ is the growth rate of structure ({again $f_{g}$ is a function of the conformal time only}), and  $G_2$ and $S_2$ is the kernel of the velocity potential. We assume that these kernels are well described by
\begin{eqnarray}
F_2({\k}_1,{\k}_2)&=&\frac{10}{7}+\frac{{\k}_1 \cdot {\k}_2}{k_1
k_2}\left(\frac{k_1}{k_2}+\frac{k_2}{k_1}\right)+\frac{4}{7}
\left(\frac{{\k}_1 \cdot {\k}_2}{k_1 k_2}\right)^2\,,
\\
G_2({\k}_1,{\k}_2)&=& \frac{6}{7} +\frac{{\k}_1\cdot {\k}_2}{k_1k_2}\left(\frac{k_1}{k_2}+\frac{k_2}{k_1}\right) +\frac{8}{7} \left(\frac{{\k}_1\cdot {\k}_2}{k_1 k_2}\right)^2\,,
\\
S_2({\k}_1,{\k}_2)& =& \frac{\left({
 \k}_1\cdot {\k}_2\right)^2}{\left(k_1 k_2\right)^2} - \frac{1}{3}\,.
\end{eqnarray}

Given equations \eqref{eq:galaxyoverfnl1} and \eqref{eq:galaxyoverfnl2}, the conformal time derivative of $v_{g}$ becomes
\begin{eqnarray}
{v\one}'_{g}({\k},\eta) &=& \frac{\HH^2 f_g}{k^2} X_{g1} \delta_{m}({\k,\eta}) \,,
\\
{v_g\two}'({\k},\eta)&=&\frac{\HH^2 f_{g}X_{g2}}{k^2} \int \frac{d^3k_1}{(2\pi)^3}\frac{d^3k_2}{(2\pi)^3}{G_2} ({\k}_{1}, {\k}_{2})
\delta_{m}({\k}_1, \eta)\delta_{m}({\k}_2, \eta)  \delta^{(3)}\left({\k} -{\k}_1 -{\k}_2\right)\,.
\end{eqnarray}
{Note that $ \bar{X}_{g1} = \HH X_{g1}$ and  $ \bar{X}_{g2} = \HH X_{g2}$,} where 
\begin{eqnarray}
X_{g1} &=&  \frac{{f_{g}}'}{\HH f_{g}} + f_{g} + \frac{\HH'}{\HH^2}\,, \qquad
X_{g2} =  \frac{{f_{g}}'}{\HH f_{g}} + 2f_{g} + \frac{\HH'}{\HH^2}\,.
\end{eqnarray}
We will assume that $\delta_g$ and $v_g$ are precisely determined by the even multipoles.
With these tools we expand equations \eqref{eq:linearordergalaxydensityKaiser} and \eqref{eq:Newtoniannumbercount2} in Fourier space. In the Newtonian limit, the Fourier space kernel becomes \cite{Bernardeau:2001qr}
\begin{eqnarray}
\mathcal{K}_{\rm{N}}\one(k_1)&=& b_1+f_{g} \mu^2_1\,, 
\\
\mathcal{K}_{\rm{N}}\two(\bm{k}_{1}, \bm{k}_{2}, \bm{k}_{3})
&=& b_{2}+ b_{1}{F}_{2}({\k}_{1},{\k}_{2})+ 
b_{s^2}S_{2}({\k}_{1},{\k}_{2})
+ f_{g}\,{G}_{2}({\k}_{1},{\k}_{2})
\mu_3^{2}
\\ \nonumber&&
 + b_1f_{g}\left[\left(\mu_1^2 + \mu_2^2\right)+\mu_1\mu_2  \left(\frac{k_1}{k_2}
+ \frac{k_2}{k_1}\right) \right]
+  f_{g}^2\left[ 2 \mu_1^2 \mu_2^2
 +{\mu_1\mu_2}\left(\mu_1^2 \frac{k_1}{k_2}+ \mu_2^2\frac{k_2}{k_1}\right) \right]
\,,
\end{eqnarray}
where $\mu_{m} = \hat{\k}_{m}\cdot {\n} = {\k}_{m}\cdot{\n}/k_{m}$ with $m \in \{1,2,3\}$. The Fourier space kernels for the Doppler part (i.e. equations \eqref{eq:linearordergalaxydensityDoppler2} and \eqref{eq:Secondordergalaxydensity2}) become
\begin{eqnarray}
\mathcal{K}_{\rm{D}}\one(k_1) &=&\HH
\frac{ f_{g} B_{1}}{k} \mu_1\, ,
 \\  \nonumber 
\mathcal{K}_{\rm{D}}\two(\bm{k}_{1}, \bm{k}_{2}, \bm{k}_{3})&=&
\HH \bigg\{\frac{f_{g}\mu_3}{k_3} A_0{G}_2({\k}_1,{\k}_2) -b_1 f_{g}A_{1} \left( \frac{\mu_1k_2^3 + \mu_1 k_1^3}{k_1^2 k_2^2}\right)
 + f_{g}A_2\left(\frac{\mu_{1}}{k_1} + \frac{\mu_{2}}{k_{2}}\right)
  \\ \nonumber &&
 + f_{g} ^2 \bigg[ A_4
\left(\bm{k}_{1}\cdot\bm{k}_{2}\right)\left(\frac{\mu_{1}k_{1} + \mu_{2}k_{2}}{k_{1}^{2}k_{2}^{2}}\right)
-\left(1+A_3\right)\left(\frac{\mu_{1}\mu_{2}^{2}k_{1}k_{2}^{2} + \mu_{2}\mu_{1}^{2}k_{2}k_{1}^{2}}{k_{1}^{2}k_{2}^{2}}\right)
    \\ &&
-A_1 \left(\frac{\mu_2^3 k_2^3 + \mu_1^3 k_1^3}{k_1^2k_2^2}\right)\bigg]
\bigg\} \,,
\end{eqnarray}
where $B_1, A_0, A_1, A_2, A_3$ and $A_4$ are redshift dependent terms
\begin{eqnarray}
B_1&=&  
   \Upsilon_1
      +\beta_{1} X_{g1}  + b_e   - 2 \mathcal{Q}  - \frac{2\left(1- \mathcal{Q}\right)}{\chi \HH} - \frac{\HH'}{\HH^2}  \,,
\\
A_0&=& \Upsilon_{2}
      + \beta_{2} X_{g2}  + b_e   - 2 \mathcal{Q}  - \frac{2\left(1- \mathcal{Q}\right)}{\chi \HH} - \frac{\HH'}{\HH^2}  \,,
\\
A_1&=& (1-\beta_{1})X_{g1}+ (1-\Upsilon_1) \,,
 \\
 A_2&=& 2\left( 1 - \frac{1}{\chi \HH}\right) \frac{\partial b_1}{\partial \ln L} 
        +b_1f\left[ \beta_{1} X_{g1}  - (1-\Upsilon_1)  \right]\,
    \nonumber \\ &&
+\left( b_{1}  f_{g} + \frac{b'_{1}}{\HH}\right)+b_{1}  \left[1+ b_e   - 2 \mathcal{Q}  - \frac{2\left(1- \mathcal{Q}\right)}{\chi \HH} - \frac{\HH'}{\HH^2} \right]\,,
    \\
A_3&=&-2(1+\beta_{1})X_{g1} +
(1+\Upsilon_1)   -2\left[ \beta_{1}X_{g1}  - (1-\Upsilon_1)  \right]\,
   \nonumber \\ &&
-2- 2b_e   + 4 \mathcal{Q}  + \frac{4\left(1- \mathcal{Q}\right)}{\chi \HH} + 3\frac{\HH'}{\HH^2}
 \\
 A_4&=&1+(1-\Upsilon_1)^2\,.
\end{eqnarray}
For the standard treatment at second order,  we set  $\Upsilon_{1,2,} =0$ and $\beta_{1,2}  = 0$. In this limit, we can make use of the Poisson equation to relate the gravitational potential to the matter density
\begin{eqnarray}
 X_{g1} = \frac{f'_{g}}{f_{g}\HH} + f_{g} + \frac{\HH'}{\HH^2} = \frac{3}{2} \frac{\Omega_m}{f_{g}} -1  \,,
\end{eqnarray}
where $\Omega_m$ is the matter density parameter, then we recover exactly the result first derived in \cite{Clarkson:2018dwn} and discussed in more detail in \cite{Maartens:2019yhx}.

\subsection{Odd multipoles of the galaxy power spectrum }\label{sec:multipolespowerspectrum}
We obtain the galaxy power spectrum, $ P_{g}^{AB}(k)$, as the expectation value of a two-point correlation function of the  galaxy number count of two dissimilar tracers $A$ and $B$  in Fourier space:
\begin{eqnarray}
P_{g}^{\rm{AB}}(k,\mu) &=&\left[\mathcal{K}^{\rm{A}}_{\rm{N}}({k},\mu)+i \mathcal{K}^{\rm{A}}_{\rm{D}}({k},\mu)\right]
\left[\mathcal{K}^{\rm{B}}_{\rm{N}}({k},-\mu)+ i\mathcal{K}^{\rm{B}}_{\rm{D}}({k},-\mu)\right]
P_{m}(z,k)\,.
\end{eqnarray}
Here $P_{g}^{\rm{AB}}$ is a complex function and $P_{m}$  is the matter power spectrum, which is real. In the weak field limit, the real part of $P_{g}^{\rm{AB}}$  corresponds to the standard Newtonian approximation or the Kaiser limit, while the imaginary part  corresponds to the Doppler part:
$
P_{g}^{AB} = P^{AB}_{g{\rm{N}}} +  i P^{AB}_{g{\rm{D}}  }
$
We can isolate the Newtonian and the Doppler parts using
\begin{eqnarray}
P^{AB}_{g{\rm{N}}} (k,\mu)  &=& \frac{1}{2} \left[P_{g}^{AB}(k,\mu)  + P_{g}^{AB}(k,\mu) ^{*}\right] \,, 
\\
P^{AB}_{g{\rm{D}}  }(k,\mu) &=& \frac{1}{2 i }\left[P_{g}^{AB}(k,\mu)  - P_{g}^{AB}(k,\mu) ^{*}\right]\,.
\end{eqnarray}
The real part is symmetric with respect to the exchange of the line of sight direction: $P^{AB}_{g{\rm{N}}} (k,\mu)   = P^{AB}_{g{\rm{N}}} (k,-\mu)  $, while the imaginary part is anti-symmetric with respect to the exchange of the  line of sight direction: $ P^{AB}_{g{\rm{D}}  }  (k,\mu) = - P^{AB}_{g{\rm{D}}  } (k,-\mu) $. Similarly, the real part is symmetric with the exchange of tracer positions: $P^{AB}_{g{\rm{N}}}  = P^{BA}_{g{\rm{N}}} $, while the imaginary part is anti-symmetric with the exchange of tracers $ P^{AB}_{g{\rm{D}}  } = - P^{BA}_{g{\rm{D}}}$. {Thus, a simultaneous exchange of tracers and the line of sight direction leaves the imaginary part unchanged.}

Tracers with dissimilar astrophysical bias parameters lead to a complex cross-power spectrum and it is possible to  expand it using  Legendre polynomials, $\mathcal{L}_{\ell}(\mu)$:  
\begin{equation}
P^{AB}_{g} (z,k,\mu) = \sum_{\ell = 0} P^{AB}_{\ell}(z,k) \mathcal{L}_{\ell}(\mu)\,.
\end{equation}
 The imaginary part of $P_{g}^{AB}$ induces  only the odd multipole moments, which are obtained by using the orthogonality property of the Legendre polynomial
\begin{eqnarray}\label{eq:multipolesgalpowerspectrum}
P^{AB}_{\ell}(k) = \frac{(2\ell+1)}{2} \int_{-1}^{1}\d\mu\, P^{AB}(k,\mu) \mathcal{L}_{\ell}(\mu) \,,
\end{eqnarray}
where the multipole moments are given in terms of the matter power spectrum as
\begin{eqnarray}
P^{AB}_0(k) &=& \left[ b^{A}_{1}b^{B}_{1} + \frac{1}{3}(b^{A}_{1}f_g^{B} + b^{B}_{1}f_{g}^{B}) + \frac{f_{g}^{A} f^{B}_{g}}{5}\right]P_m(k) \,, \\ 
P^{AB}_1(k) &=& (-i)\bigg[\left(b^{A}_{1}b_e^{B} f_{g}^{B}-b^{B}_{1}b_e^{A}f^{A}_{g}  \right)
+(b^{B}_{1}f^{A}_{g} - b^{A}_{1}f^{B}_{g})  \bigg[\frac{2}{\chi\mathcal{H}} + \frac{{\mathcal{H}'}}{\mathcal{H}^2}- \Upsilon_1      -\beta_{1} X_{g1}\bigg]
\\  \nonumber &&
 +{f^{A}_{g}}{f^{B}_{g}}\bigg[ \frac{3}{5}  \left( b_{e}^{B}- b_{e}^{A}\right)
+ 3\left(1 - \frac{1}{\chi\mathcal{H}}\right)(s^A - s^B) \bigg]  
\nonumber \\ &&
+ 5\left(1 - \frac{1}{\chi\mathcal{H}}\right)\left( b^{B}_{1}s^A f^{A}_{g}- b^{A}_{1}s^Bf^{B}_{g}\right)
\bigg] \frac{\mathcal{H}}{k} P_m(k) \,,\\ 
P^{AB}_2(k) &=&\left[\frac{2}{3}(b^{A}_{1}f^{B}_{g} + b^{B}_{1}f^{A}_{g}) + \frac{4f^{A}_{g}f^{B}_{g}}{7}\right]P_m(k) \,, \\
P^{AB}_3(k) &=& 2i\bigg[\frac{1}{5} (b_{e}^{A}-b_{e}^{B})-\left(1 - \frac{1}{\chi\mathcal{H}}\right)(s^A- s^B)\bigg]
f_{g}^{A}f_{g}^{B}\frac{\mathcal{H}}{k}P_m(k)\,,\\
P^{AB}_4(k) &=& \frac{8}{35}f_{g}^{A}f_{g}^{B}P_m(k),
\label{eq:powerspectrummults}
\end{eqnarray}
%
{where we have assigned each tracer with a corresponding growth rate for completeness.  We note that the baryon-to-dark matter ratio is very similar in different galaxies, therefore we assume it is determined by the cosmological background value as is done in \cite{Gleyzes:2015rua}, hence,  we set $ f_{g}^{A} =f_{g}^{B} =f_{g}$ in  the subsequent quantitive analysis. } The odd multipoles vanish in the limit where $A =B$~\cite{Hall:2016bmm}.

\subsection{Odd multipoles of the galaxy Bispectrum}\label{sec:multipolesbispectrum}

Contrary to the galaxy power spectrum case,  the galaxy bispectrum for a single tracer in the weak-field limit is a complex function 
\begin{eqnarray} \label{eq:bispectrumdef}
B_{g}( z,{\k}_{1},  {\k}_{2},  {\k}_{3})   \equiv B_{g}^{{\rm{N}}}( z,{\k}_{1},  {\k}_{2},  {\k}_{3}) + iB_{g}^{{\rm{D}}}( z,{\k}_{1},  {\k}_{2},  {\k}_{3}) \,,
\end{eqnarray} 
where we have identified the real part with the Newtonian limit of the galaxy bispectrum, $B_{\rm{N}}$; it is given by~\cite{Scoccimarro:1999ed}
\begin{eqnarray}\label{eq:bispectrumg}
B_{g}^{{\rm{N}}}( {\k}_{1},  {\k}_{2},  {\k}_{3}) = \mathcal{K}_{\rm{N}}\one({\k}_{1}) \mathcal{K}_{{\rm{N}}}\one({\k}_{2}) \mathcal{K}_{\rm{N}}\two({\k}_{1},  {\k}_{2},{\k}_{3}) P_{m}(k_{1})P_{m}(k_{2}) +   \text{2 cy. p.} 
\end{eqnarray}
The imaginary part corresponds to the galaxy bispectrum induced by the Doppler effects, and its leading contribution is given by~\cite{Umeh:2016nuh,Clarkson:2018dwn}
\begin{eqnarray}\label{eq:Dopbispectrum}
B_{g}^{{\rm D}}( {\k}_{1},  {\k}_{2},  {\k}_3) &=& \Big\{
{\mathcal{K}}_{\rm N}\one({\k}_1){\mathcal{K}}_{\rm N}\one({\k}_2)\mathcal{K}_{\rm D}\two({\k}_1,{\k}_2,{\k}_3)
\nonumber\\
&&~~~ 
+\Big[ {\mathcal{K}}_{\rm N}\one({\k}_1)\mathcal{K}_{\rm D}\one({\k}_2)+{\mathcal{K}}_{\rm D}\one({\k}_1)\mathcal{K}_{\rm N}\one({\k}_2) \Big]\mathcal{K}_{\rm N}\two({\k}_1,{\k}_2,{\k}_3)
\Big\}
\nonumber\\ && \qquad \times
P_{m}(k_{1})P_{m}(k_{2}) +\text{2 cy. p.} 
\end{eqnarray}
Without loss of generality, we work in the plane-parallel limit and express all the three angles in terms of $\mu_1 = {\hat{\k}_1 \cdot{\n}}$. Requiring that ${\k}_1, {\k}_2$ and ${\k}_3$ form a closed triangle, we find
\begin{eqnarray}\label{eq:angles}
\mu_{2} &=& \mu_{1}\mu_{12} \pm \sqrt{1 - \mu_{1}^{2} }\sqrt{1 - \mu_{12}^{2}}\cos \phi_n\,,
\\
\mu_3 &=& - \frac{k_1}{k_3} \mu_1 - \frac{k_2}{k_3} \mu_2\,,
\end{eqnarray}
where $\mu_{12} = {\hat{\k}_1}\cdot{\hat{\k}_2}$. In this limit, the number of parameters reduces to five, i.e.  equations \eqref{eq:bispectrumg} and \eqref{eq:Dopbispectrum} depend only on  five parameters $B_g({k}_1,{k}_2,\mu_{12},\mu_{1},\phi_n)$. The angular dependence of the galaxy bispectrum relative to the line of sight, i.e  $\mu_1$ and $\phi_n$ may be expanded in spherical harmonics $Y_{L M}$
 \cite{Scoccimarro:1999ed,Scoccimarro:2000sn},
\begin{eqnarray}\label{eq:Bgmultipoles}
B_g({k}_1,{k}_2,\mu_{12},\mu_{1},\phi_n) &=& \sum_{L = 0}^{\infty}
\sum_{M = -L}^{L}B_{gL M} ({k}_1,{k}_2,\mu_{12}) Y_{L }^{M}(\mu_1,\phi_{n}), 
\end{eqnarray}
where $B_{gL M}$ is the multipole moments of the galaxy bispectrum
\begin{eqnarray}\label{eq:Bmultipoles1}
B_{gL M}({k}_1,{k}_2,\mu_{12}) &=&  \int_{-1}^{1} \d \mu_1\,\int_{0}^{2\pi} \d \phi_n
B_g({k}_1,{k}_2,\mu_{12},\mu_1,\phi_n) {Y_{L }^{M}}^{*}(\mu_1,\phi_{n})\,.
\end{eqnarray}
For simplicity, we shall consider only the $M=0$ moments which reduce to an azimuthal angle averaged galaxy bispectrum or the $\phi_{n}$-average multipole moments  of the galaxy bispectrum 
\begin{eqnarray}
B_{g }^{\phi_n}({k}_1,{k}_2,\mu_{12},\mu_1) \equiv \int_{0}^{2\pi} 
B_g({k}_1,{k}_2,\mu_{12},\mu_1,\phi_n)\frac{\d \phi_n}{2\pi} \,.
\end{eqnarray}
Averaging over the azimuthal angles helps to improve the signal to noise ratio \cite{Bartolo:2004if}. Also, it was shown in \cite{Gagrani:2016rfy} that the information loss associated with averaging over the azimuthal  angle  (or setting $M = 0$) is negligible. We  are interested in the multipoles of azimuthal angle averaged galaxy bispectrum, which we compute as
\begin{eqnarray}\label{eq:multipolesofBg}
B_{g{L}}({k}_1,{k}_2,\mu_{12}) &=& {{(2L+1)\over 2}}
  \int_{-1}^{1}\d\mu_{1}\, \int \frac{\d \phi_n}{2\pi} B_g({k}_1,{k}_2,\mu_{12},\mu_{1}) {\cal{L}}_{L } ( \mu_1 ) .
\end{eqnarray}
The even multipoles are sourced by the  Newtonian galaxy bispectrum (equation \eqref{eq:bispectrumdef}) while the odd multipoles are sourced by the Doppler galaxy bispectrum (equation \eqref{eq:Dopbispectrum}).  {The full  multipole decomposition of the relativistic galaxy bispectrum is given in~\cite{deWeerd:2019cae}.}

{Finally, we made a plane-parallel approximation to show only the odd multipoles of the cross-power spectrum and the bispectrum that are induced by the general relativistic effects. At the power spectrum level, the wide-angle corrections from the standard density and redshift-space distortion term generate odd multipoles with the same $\HH/k$  scale dependence as well. This additional contribution  is further suppressed by the ratio of the separation between the sources to the distance to the source.  This contribution is obviously sub-dominant in  plane-parallel approximation but would become important for a survey that covers the full sky~\cite{Bonvin:2013ogt}.  For the bispectrum, there has not yet been a study of this in detail; however, we expect that the wide-angle corrections from the second-order density and redshift-space distortion terms will lead to a similar conclusion as in the cross-power spectrum case. }

\section{Fisher forecast analysis of stage IV spectroscopic survey}\label{sec:Fisherforecast}

Our plan is to ascertain how well odd multipole moments of the galaxy cross-power spectrum and bispectrum could constrain the equivalence principle violation through the measurement of these parameters given a stage IV spectroscopic galaxy survey; $\{\Upsilon_1, \Upsilon_{2}, \beta_{1},  \beta_{2}\}$.
These parameters are zero if the galaxy motion obeys the equivalence principle, hence any non-zero detection will be an indicator of the violation of the equivalence principle on cosmological scales.

We focus on the late-time violation of the equivalence principle and parametrise $\Upsilon_{1,2}$ as
\begin{eqnarray}\label{eq:Eqparametrization}
\Upsilon_{1,2}(z) = \frac{1 -\Omega_m(z)}{1-\Omega_{m} }\gamma_{1,2}\,, 
\end{eqnarray}
where we have assumed a redshift dependence suggested by \cite{Bonvin:2018ckp}.
This fixes $\beta_{1,2}$ using equation \eqref{eq:fixbeta12}, which implies that we are left with only two redshift-independent parameters to describe the violation of the equivalence violation:
\begin{eqnarray}
{\rm{Equivalence~principle~violation~parameters}}  &= &  \left\{\gamma_1, \gamma_{2}\right\}\,.
\end{eqnarray}
In order to optimise constraints on these parameters, we fix the following cosmological parameters~\cite{Aghanim:2018eyx}:
\begin{eqnarray}
{\rm{Cosmological~ parameters }} &=& \left\{ D_m, f_g, X_{g1}, X_{g2}, P_{\cal O}(k)\right\}\,.
\end{eqnarray}
{These parameters will be well constrained by the combination of the Cosmic Microwave Background (CMB) anisotropies and even multipole moments of the power spectrum and bispectrum as well as weak gravitational lensing.
}
Similarly,  we assume that the parameters that characterise the  Alcock Paczynski effect~\cite{1979Natur.281..358A} can be fully determined by the even multipoles of the N-point correlation function~ \cite{Nadathur:2019mct,deMattia:2020fkb}, hence we neglect these effects in the subsequent analysis. 
In addition to the cosmological parameters, we have to determine the following astrophysical parameters:
\begin{eqnarray}
{\rm{Astrophysical~parameters}} = \left\{b_1, b_2,  b_{s^2} ,b_e, \mathcal{Q},\frac{\partial b_1}{\partial \ln{\bar{L}}}\right\}\,.
\end{eqnarray}
For these parameters  we consider  a Stage IV $\Ha$ emission line galaxy  survey and HI intensity mapping survey. We focus on this combination of tracers because an earlier study \cite{Hall:2016bmm} has shown that the two-point correlation function of the HI intensity mapping and H$\alpha$-emission line galaxy spectroscopic galaxy survey leads to a high signal-to-noise ratio (SNR) for detecting the dipole moment.

\begin{itemize}
\item {\tt{ $\Ha$ emission line galaxies}}

We consider a distribution of the H$\alpha$ emission line galaxies with a luminosity function \cite{Pozzetti:2016cch}
\begin{equation}\label{eq:Halum}
\Phi_L(z,L) =  \frac{\bar{\Phi}_{\star}(z)}{\bar{L}_{\star}(z)} \left(\frac{L}{\bar{L}_{\star}(z)}\right)^{\alpha} \exp\left(-\frac{L}{\bar{L}_{\star}(z)}\right)   \,,
\end{equation}
where  $\bar{L}_{\star}(z)$ is the critical luminosity at a given redshift and is assumed to evolve as  $\bar{L}_{\star}(z)  = \bar{L}_{\star,0}  \left(1 + z\right)^{2 }$. Here  $L_{\star,0}$ is the  critical luminosity today with the best-fit value $\log L_{\star,0}={41.50}$ erg s$^{-1}$, $\alpha$ is the faint-end slope with the best fit value $\alpha= -1.35$,  and  $\bar{\Phi}_{\star}(z)$  is the characteristic number density at $z$
\begin{eqnarray}
\bar{\Phi}_{\star}(z)=
\begin{cases}
\bar{\Phi}_{\star,0}  \left(1+z\right)  & \text{for} \quad z<z_{\rm break} \,,\\ \\
\bar{\Phi}_{\star,0} \left(1+z_{\rm break}\right)^{2}/ \left(1+z\right) &\text{for} \quad z>z_{\rm break}\,.
\end{cases}
\end{eqnarray}
$\Phi_{\star,0}$ is the characteristic number density today with the best fit value  $\log \bar{\Phi}_{\star,0}={-2.8}$ Mpc$^{-3}$  and $z_{\rm break}=1.3$.    The evolution and magnification  bias  parameters  are obtained from equation \eqref{eq:Halum} \cite{Maartens:2019yhx}. 
The number density of  the H$\alpha$ emission line galaxy is obtained from the luminosity function
\begin{eqnarray}
 n_{g}^{\Ha}(z) = \int_{\ln L} \Phi_L(z,L) \d \ln L\,.
\end{eqnarray} 
For the  clustering bias parameters we use the values given  in  \cite{Yankelevich:2018uaz}
\begin{eqnarray}\label{eq:Halpha1}
 b^{\Ha}_{1} \left( z\right)&=& 0.9 + 0.4 z \, ,
 \\
 b^{\Ha}_{2}\left(z\right)&=& -0.704172 -0.207993 z +0.183023 z^{2}-0.00771288 z^3 \, ,
 \label{eq:Halpha2}
 \\
   b^{\Ha}_{s^2}\left(z\right)&=&0.0321163-0.123159 z +0.00694159 z^2-0.00171397 z^3\,.  
  \label{eq:Halphatidal}
\end{eqnarray}
Finally, we set 
\begin{eqnarray}
\frac{\partial b_1^{\Ha}(z,L)}{\partial \ln L} = 0
\end{eqnarray}
since $b_1^{\Ha}(z,L)~$ is nearly constant in $L$\cite{Maartens:2019yhx}.

\begin{figure}[h]
\centering 
\includegraphics[width=150mm,height=60mm] {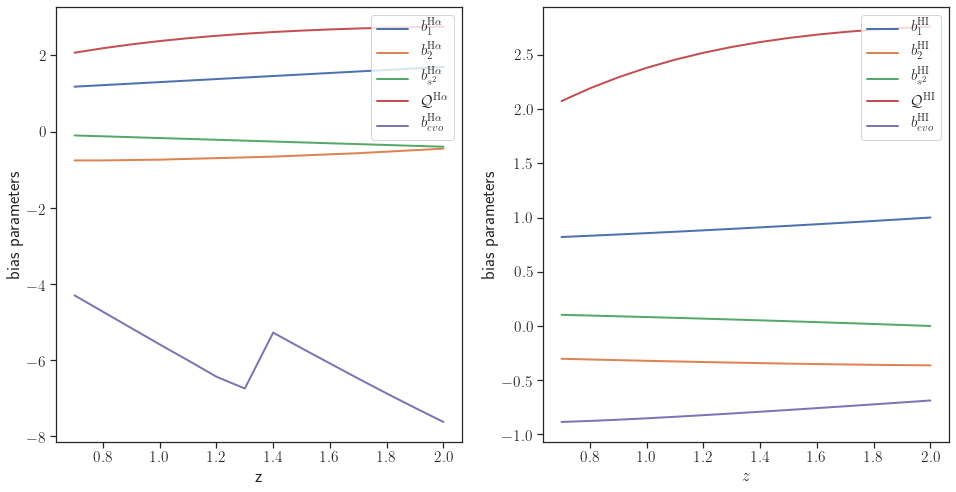}
\caption{\label{fig:BiasEv} Redshift evolutions of the bias parameters. The left panel shows bias parameters for the H-alpha survey while the right panel shows them for the HI survey.  The feature in the H-alpha evolution bias corresponds to the peak of the luminosity function at $z_{\rm break}= 1.3$.}
\end{figure}
\item {\tt{HI intensity mapping survey}}

For the HI intensity mapping survey, we  follow the  prescription given in \cite{Santos:2015gra,Umeh:2015gza,Umeh:2016thy} on how to estimate the HI bias parameters from the halo bias parameter. Under this framework, the  local HI bias parameters are obtained  by assuming that the HI sources are found in galaxies which are resident in halos within a given range of circular velocities. The modelling of this leads to the following HI bias parameters~\cite{Umeh:2016thy};
\begin{eqnarray}
b^{\HI}_1(z) &=&  0.750 + 0.087 z+ 0.019 z^2\,,
\\
b^{\HI}_2(z) &=& -0.257- 0.063 z - 0.007 z^2 + 0.006 z^3\,,
\\
b^{\HI}_{s^2}(z) &=& -\frac{4}{7} \left(b^{\HI}_1(z) -1\right) \,.
\end{eqnarray}
The  mean HI brightness temperature $\bar{T}$ on the FLRW background  is given by \cite{Santos:2015gra}
  \begin{eqnarray}\label{eq:backgdeltaTbin}
  \bar{T}_{\HI} (z) &=&\frac{3\pi^2}{4} \frac{\hbar^3 A_{10}}{k_{B}E_{21}}\,
\frac{ \bar{n}_{\HI} (z)a(z)^3}{\HH(z)}\approx  566 h \frac{\Omega_{\HI}(z)}{0.003}(1+z)^2\frac{H_0}{\HH(z)} \,\,[\mu {\rm K}],
\end{eqnarray} 
where $\Omega_{\HI}$ is the comoving HI mass density.
 The evolution bias is given by ~\cite{Umeh:2016thy}
 \begin{equation}\label{eq:evolbias}
b^{\HI}_{\rm{e}} (z)
=-1.248 - 0.147 z + 0.145 z^2 - 0.012 z^3 \,.
     \end{equation}
HI has a constant magnification bias parameter $\mathcal{Q}^{\HI} = 1$ or $s =2/5$ \cite{DiDio:2016gpd}, therefore, $b_1^{\HI}$ is constant in luminosity
\begin{eqnarray}
\frac{\partial b_1^{\HI}(z,L)}{\partial \ln L} = 0\,.
\end{eqnarray}
Figure \ref{fig:BiasEv} shows the redshift evolution of the  bias parameters that will be used in the Fisher forecast.

\end{itemize}

\subsection{Estimators of the dipole moment  of the galaxy cross-power spectrum}

\begin{figure}[h]
\centering 
\includegraphics[width=150mm,height=70mm] 
{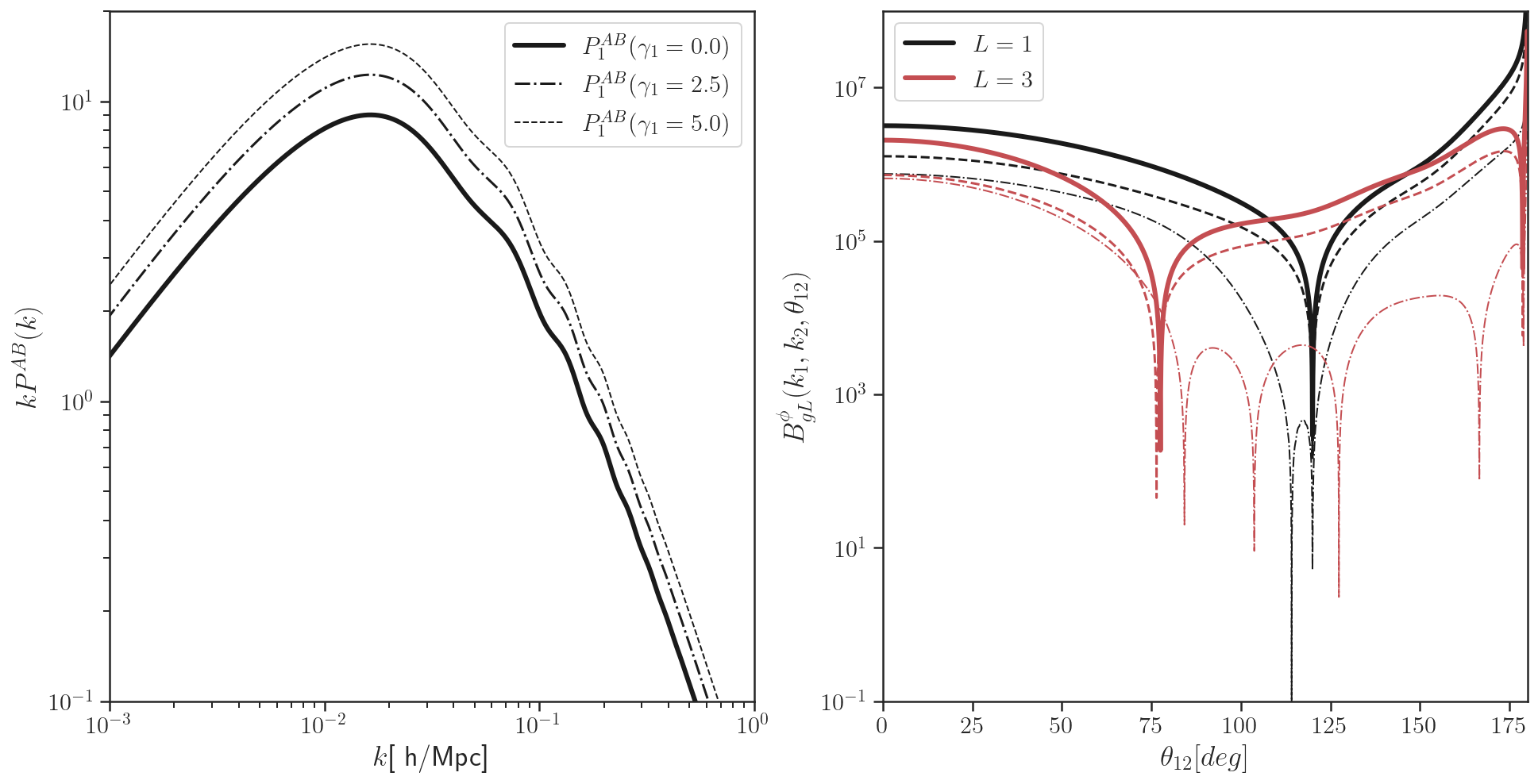}
\caption{\label{fig:powerandbispectrum} {Left panel: The thick line ($\{\gamma_1, \gamma_2\} = \{0, 0\}$)  indicates the dipole of the galaxy cross-power spectrum of  the H$\alpha$  emission line galaxy  and the HI intensity mapping  for a stage IV experiment in the limit of vanishing relative velocity between galaxies and baryons. The dashed ($\{\gamma_1, \gamma_2\} = \{2.5, 2.5\}$) and dashed-dot ($\{\gamma_1, \gamma_2\} = \{5.0,5.0\}$) lines indicate different amplitude of the relative galaxy-baryon velocity with respect to the galaxy velocity.  Right panel: The thick line indicates the bispectrum of the H$\alpha$  emission line galaxy in the limit  of vanishing galaxy-baryon relative velocity. Similarly, the dashed and dashed-dot lines indicate different amplitude of the galaxy-baryon relative velocity with respect to the galaxy velocity just as in the left panel.}
} 
\end{figure}

We define the estimator of the dipole moment of the galaxy cross-power spectrum for the dissimilar tracers as 
the band-power average of the two-point correlation function in Fourier space averaged over all lines of sight and weighted by the angle between ${\k}$ and ${\n}$
\begin{eqnarray}\label{eq:estimators}
\overline{P}^{AB}_{g1}(k_i)&\equiv & \frac{3}{2}\,
\frac{V_s}{V_{12}}\, \int_{-1}^{1} \,\d \mu\,\mu \int_{{\cal K}_i} {{\rm d}^3 k},\,\Delta_{\mathrm{g}}^{\rm{A}}({\bf{k}})\,\Delta_{\mathrm{g}}^{\rm{B}}(-{\bf{k}})\,,
\end{eqnarray}
where $V_s$ is the volume of the survey and the estimator is normalised by corresponding  volume of the k-bin ${{\cal K}_i}$,
$
V_{12} \simeq {4\pi}k_i^2 \Delta k.
$ 
The estimator defined in equation \eqref{eq:estimators} for discrete tracers is related to the theory (continuous) galaxy cross-power spectrum according to  \cite{Smith:2008ut}
\begin{eqnarray}\label{eq:powerspectrumnoise}
\< \Delta^{A}_{g}({\k}_1)\Delta^{B}_{g}({\k}_2)\> V_s  =  \hat{P}^{AB}_{g}({\k}_1)\delta^{K}_{{\k}_1,- {\k}_2}= P^{AB}_{g}({\k}_1) \delta^{K}_{{\k}_1,- {\k}_2}+ \frac{\delta_{AB}^{K}}{\bar{n}_{g}^{A}}\delta^{K}_{{\k}_1,- {\k}_2} \,,
\end{eqnarray}
where $\delta_{k_1, k_2}$ is the Kronecker delta and $P^{AB}_{g}$ is the theory (continuous) cross-power spectrum shown in Figure \ref{fig:powerandbispectrum}. 

%
In the continuum limit, it becomes
\begin{eqnarray}
\delta^{K}_{{\k}_1,{\k}'_1} \rightarrow \frac{1}{V_s}\delta^{D}\left({\k}_{1}  - {\k}_{2}\right) (2\pi)^{3}  =  \frac{1}{V_s}\frac{1}{k^2}\delta\left( {k} - {k}'\right) \delta \left({\k}_{\bot}-{\k}'_{\bot}\right)(2\pi)^{3} \,.
\end{eqnarray}
To  obtain  the covariance of the dipole of the cross-power spectrum,  we average over the angular $k$-directions (${\rm d}^3 k = k^2 \d k \d^2 \hat{\k}$)  to obtain an estimator that depends on the wavenumber
\begin{eqnarray}
{\rm{Cov}}\left[\overline{P}^{AB}_{g1}(k_i) \overline{P}^{AB}_{g1}(k_j)\right]
&=&\frac{9}{4}
\frac{1}{V_{12}}\frac{1}{V'_{12}} \int {\d k k^2}\,
\int{ \d k' k^2}  \int{{\rm d}^2 \k}\int {{\rm d}^2 \k'}
\\ \nonumber && \times
\int { d {\mu}}\int { d {\mu}'}  \mu \mu'
 {\rm{Cov}}\left[\hat{P}^{AB}_g({\k}) 
\hat{P}^{AB}_g({\k}') \right] \,.
\end{eqnarray}
The covariant matrix for the galaxy cross-power spectrum is given by \cite{Smith:2008ut}
\begin{eqnarray}
{\rm Cov}\left[\hat{P}^{AB}_g({\k}) \hat{P}^{AB}_g({\k}')\right] 
&\equiv&  
 \frac{(2\pi)^{3} }{V_s}\bigg[\hat{P}^{AA}_{g}({k},\mu)\hat{P}^{BB}_{g}({k}',-\mu')\delta^{D}
\left({\k}  + {\k'}\right) 
\\ \nonumber && \qquad	  \qquad
+  \hat{P}^{AB}_{g}({k},\mu) \hat{P}^{BA}_g({k}',-\mu')\delta^{D}\left({\k}  - {\k'}\right)\bigg]\,.
\end{eqnarray}
Note that  $\hat{P}^{BA}_{g}(k,-\mu) = \hat{P}^{AB}_{g}(k,\mu)$ and  $P_{BB}(k,-\mu) = P_{BB}(k,\mu)$. We  expand the angular dependence of the galaxy power spectrum in Legendre polynomials:
$\hat{P}^{AB}_{g}(k,\mu) =\sum_{L = 0}^{4}\hat{P}^{AB}_{gL}(k) \mathcal{L}_{L}(\mu)\,.$
We find that the multipole moment is given by
\begin{eqnarray}
\hat{P}^{AB}_{gL}(k) &=& \frac{(2L+1)}{2} \int_{-1}^{1}\d\mu\, \hat{P}^{AB}_{g}(k,\mu)\mathcal{L}_{L}(\mu)  = {P}^{AB}_{gL}(k)  + \delta_{AB}^{K}\delta_{{L0}}P^{AA}_{\rm{noise}}\,,
\end{eqnarray}
where $P^{AA}_{\rm{noise}}$ is the noise power spectrum. %
Expanding $\hat{P}^{BA}_{g}(k,-\mu) $ in terms of the Legendre polynomial gives
$$\hat{P}^{BA}_{g}(k,-\mu)=\sum_{L = 0}^{4}\hat{P}^{BA}_{gL}(k) \mathcal{L}_{L}(-{\k}\cdot{\n}) = \sum_{L = 0}^{4}(-1)^{L}\hat{P}^{BA}_{gL}(k) \mathcal{L}_{L}({\k}\cdot{\n})\,,$$
where we made use  of  the parity transformation property of the Legendre polynomial $\mathcal{L}_{L}(-{\k}\cdot{\n}) = (-1)^{L}\mathcal{L}_{L}({\k}\cdot{\n})$. Note also that $\hat{P}^{BA}_{gL}(k)  =(-1)^L \hat{P}^{AB}_{gL}(k) $. Putting all these together leads to 
\begin{eqnarray}\nonumber 
{\rm{Cov}}\left[\overline{P}^{AB}_{g1}(k_i) \overline{P}^{AB}_{g1}(k_j)\right]&=&
\frac{\delta^{K}_{{k}_i,-{k}'_j}  }{N_{k}} \sum_{L_1L_2}\bigg[\hat{P}^{AA}_{g L_1}({k}_i)\hat{P}^{BB}_{g L_2}({k}'_j)
- \hat{P}^{AB}_{g L_1 }({k}_i) \hat{P}^{AB}_{g L_2}({k}'_j)\bigg] 
\\  && \times
\int { d \mu}
\mu^2
 \mathcal{L}_{L_1}(\mu)\  \mathcal{L}_{L_2}(\mu) \,,
 \label{eq:dipolescovariance}
\end{eqnarray}
where we defined $N_{k} \equiv {4\pi k^2\Delta k V_s} /{(2\pi)^{3} }$ and made use of the relationship between a 1D Dirac delta function and the Kronecker delta 
$\delta^{D}\left({k} + {k}'\right) \rightarrow \delta^{K}_{{k}_1,-{k}'_1}/{\Delta k}  $. The minus sign in the second equality comes from performing the angular integral over the delta function. We perform the integral over $\mu$ in equation \eqref{eq:dipolescovariance} analytically and summed $\{L_1,L_2\}$  up to $\{4,4\}$  to find 
\begin{eqnarray}\nonumber
{\rm{Cov}}\left[\overline{P}^{AB}_{g1}(k_i) \overline{P}^{AB}_{g1}(k_j)\right]&=&\frac{\delta^{K}_{{k}_i,-{k}'_j} 
 }{N_{k}}
 \bigg\{\frac{3}{2}\bigg[{P}^{AA}_{g 0}({k}_i)  {P}^{BB}_{\rm{noise}} +  {P}^{BB}_{g 0}({k}_j) {P}^{AA}_{\rm{noise}  }
 +  {P}^{AA}_{\rm{noise} }  {P}^{BB}_{\rm{noise}} \bigg]
  \\ \nonumber &&
+
\frac{3}{5}\bigg[{P}^{AA}_{\rm{noise} }{P}^{BB}_{g 2}({k}_j)+{P}^{AA}_{g 2}({k}_i){P}^{BB}_{\rm{noise}}\bigg] 
- \frac{9}{10} {P}^{AB}_{g 1 }({k}_i) {P}^{AB}_{g 1}({k}_i) 
\\  &&
-\frac{23}{70}{P}^{AB}_{g 3 }({k}_i) {P}^{AB}_{g 3}({k}_i) 
- \frac{18}{35} {P}^{AB}_{g 3 }({k}_i) {P}^{AB}_{g 1}({k}_i) 
\bigg\}\,.
\label{eq:dipolecovariance2}
\end{eqnarray}
Equation \eqref{eq:dipolecovariance2} agrees with  \cite{Beutler:2020evf}. 


\subsection{Estimators of the odd multipoles of the galaxy bispectrum}
We define the estimator for the azimuthal angle averaged multipole moments of the galaxy bispectrum following \cite{Regan:2017vgi} as
\begin{eqnarray}
\overline{B}_{g{\ell }}({k}_i,{k}_j,k_k) &\equiv & \frac{2\ell+1}{2}\frac{V_{k_f}}{V_{123}}\,\int_{{\cal T}_{i,j}} {{\rm d}^3 k_1\,{\rm d}^3 k_2 {\rm d}^3 k_3}
\int \frac{{\d^2{\n}}}{4\pi}\, \int_0^{2\pi}\frac{ \d\phi}{2\pi} \,\,
\\ \nonumber && \qquad \times
\Delta^{\rm{A}}_{\mathrm{g}}({{\k}_1})\,\Delta^{\rm{A}}_{\mathrm{g}}({\k}_2)\,\Delta^{\rm{A}}_{\mathrm{g}}({{\k}_3})
\delta^{\rm{D}}\left({\k}_1 + {\k}_2 + {\k}_3\right)
 \mathcal{L}_{\ell}({\k}_1\cdot{\n})\,,
\end{eqnarray}
where $V_k=k_f^3$ is the fundamental $k$-space cell-volume, $k_f$ is the fundamental wavenumber, which is related to the volume of the survey according to $k_f=2\pi/L = 2\pi/V_s^{1/3}$, and $V_{123}$ is the effective k-space volume of the k-bin decomposed into spherical shells
\begin{eqnarray}
V_{123}&=&\int_{{\cal T}_i}\delta^{\rm D}({\bm{p}}+{\bm{q}}+{\bm{k}})\,{\rm d}^3 p\,{\rm d}^3 q\,{\rm d}^3 k \simeq 8\pi^2 k_1 k_2 k_3 (\Delta k)^3  \beta(\mu_{12})\;.
\end{eqnarray}
Here $\beta(\mu_{12})$ is a normalisation factor that depends on the shape of the triangular configuration~\cite{Chan:2016ehg}
\begin{eqnarray}
   \beta(\mu_{12} )=
    \begin{cases}
      \frac{1}{2} & \text{if}\quad \mu_{12}  =\pm 1 \\
      1 & \text{if}\quad \ 0< \mu_{12}  < 1 \\
      0 & \text{otherwise}
    \end{cases}
\end{eqnarray}
In the Gaussian limit, the covariance of the galaxy bispectrum becomes \cite{Gagrani:2016rfy}
\begin{eqnarray}\nonumber 
{\rm{Cov}}\left[\overline{B}^A_{g\ell }\overline{B}^B_{g\ell'}\right]
&\simeq &\frac{(2\ell +1)}{2}\frac{(2\ell' +1)}{2} \delta^K_{AB} \frac{s_{\rm B}\,V_s\,}{{N}_B}
\displaystyle\int \!\mathrm{d}\mu_1\int\frac{\mathrm{d}\phi }{2\pi}
{\hat{P}^{AA}_{g}(k_1,\mu_1)\,\hat{P}^{AA}_{g}(k_2,\mu_2)\, \hat{P}^{AA}_{g}(k_3,\mu_3)} 
\\  && \times
\mathcal{L}_{\ell}(\mu_1)\mathcal{L}_{\ell'}(\mu_1) \,,
\label{eq:BispectrumGaussianlimit}
\end{eqnarray}
where  $s_B = 6,2,1$ for equilateral, isosceles and general triangles, respectively and 
\begin{equation}
{N}_B\simeq  \frac{1}{\pi}\frac{V^2}{(2\pi)^3}{k}_{ 1}\,{k}_{ 2}\,{k}_{3}\,(\Delta k)^3 \beta(\mu_{12} )\,.
  \end{equation}
Each of the power spectrum is decomposed with respect to ${\n}$:
$\hat{P}^{AA}_{g}(k_i,\mu_i) ={P}^{AA}_g(k_i,\mu_i)+{1}/{\bar{n}^{\rm{H}\alpha}_g}$. We use equation \eqref{eq:angles} to relate  $\{\mu_3, \mu_{2}\}$ to  $\mu_{1}$ and the azimuthal angle $\phi_n$.  We performed the integrals over $\mu_{1}$ and  $\phi_n$ analytically using MATHEMATICA. 


\subsection{Fisher forecasts for the equivalence principle violation constraint}

Firstly, we compute the signal to noise ratio (SNR) for the overlapping  HI and $\Ha$ spectroscopic surveys using
\begin{eqnarray}
 \left(\frac{S}{N}\right)^2 &=&\sum_{z_{\rm{min}}}^{z_{\rm{max}}}\sum_{T_{\rm{X}}}
{X_{\ell}}(k_i)
{\rm Cov}^{-1}\left[X_{\ell}(k_i),X_{\ell}(k_j)\right]
 {X}^{H}_{\ell}(k_j),
\end{eqnarray}
where ${X}^{H}_{\ell}$ is the Hermitian conjugate of ${X}_{\ell} = \{P^{AB}_{1}({k_i}), B_1(k_i,k_j,k_k),B_3(k_i,k_j,k_k)\}$, and the summation sign  is defined as 
\begin{equation}
\sum_{T_{\rm{B}}} \equiv \sum	_{k_1 = k_{\rm{min}}}^{k_{max}} \sum	_{k_2 = k_1}^{k_{11}}  \sum	_{k_3 = k_{\star}}^{k_{2}}  \qquad {\rm{and}}\qquad  \sum_{T_{\rm{P}}} = \sum_{k, k'\le k_{\rm max}}
\end{equation}
for the galaxy bispectrum  and  cross-power spectrum, respectively. 
We take the cross-covariances (power spectrum-bispectra and dipole-octupole bispectra covariances) to be zero and set $k_{\rm{max}}$ at the maximum scale below which the perturbation theory can be trusted: $k_{\rm{max}} = 0.1(1+z)^{(2/(2+ n_s))} [h{\rm{Mpc}^{-1}}]$~\cite{Blanchard:2019oqi}, $k_{\rm{min}}$ is determined by the volume of the survey $k_{\rm{min}} \sim1/V_s^{1/3}$ and to ensure that the closure property of the triangle is satisfied we have $k_{\star} = \rm{max}(k_{\rm{min}}, k_1-k_2)$. Also, we set the width of the k-bins to $\Delta k = 3 k_{\rm{min}}$. {We neglect the covariance between the power spectrum and the bispectrum since we only consider the Gaussian covariance. Also we consider only a single tracer at the bispectrum level. For a single tracer, the odd multipole moment in the power spectrum vanishes. The covariance between the dipole and the octupole of the bispectrum is neglected for simplicity, which is not expected to bias the result since the majority of the signal is contained in the dipole.}
\begin{figure}[h]
\centering 
\includegraphics[width=150mm,height=70mm] {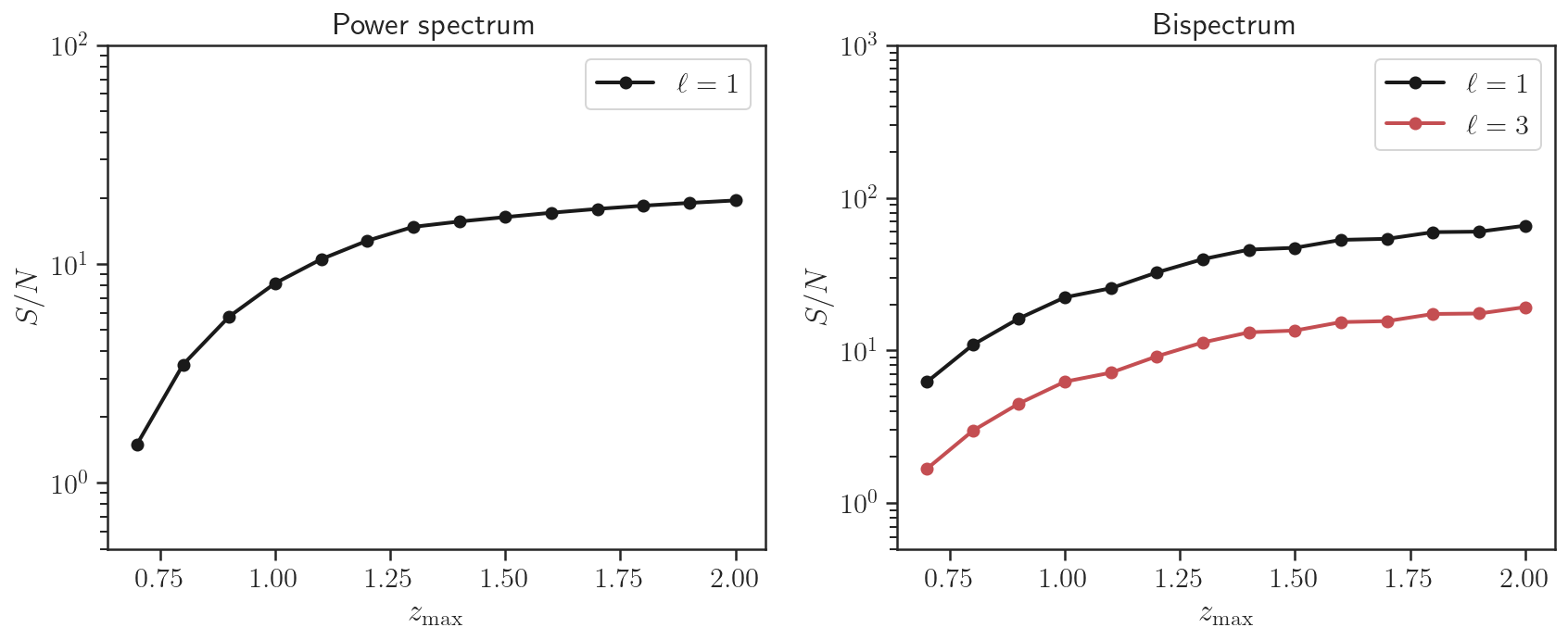}
\caption{\label{fig:SNR} {Left panel: This is the SNR for detecting the dipole of the galaxy cross-power spectrum between the H$\alpha$ line emission galaxy and the HI intensity mapping survey in the single dish mode.   Right panel: This is the SNR for detecting the dipole and octupole moments of the H$\alpha$ bispectrum.  We take the fiducial values to be $\{\gamma_1,  \gamma_{2}\} \rightarrow \{0,0\}$.}}
\end{figure}

The surveys have the following noise attributes; for the H$\alpha$ emission line galaxy survey, the noise budget is dominated by the shot noise while the HI intensity mapping survey, the noise budget is dominated by the system and sky temperatures
\begin{eqnarray}
P_{\rm{noise}}^{\Ha\Ha}&=& \frac{1}{n_{g}^{\Ha}}\,,
\qquad
P_{\rm{noise}}^{\HI \HI} =4\pi f_{\rm{sky}} \frac{\chi(z)^2 (1+z)^2 T_{\rm{sys}}(z)^2}{N_{\rm{pol}} N_{d} H(z) \nu_{21} t_{tot} T^2_{\rm{HI}}(z)} \,, \,\,\,
\label{eq:HInoice}
\end{eqnarray}
where
$N_{\rm{pol}}  =2$ in equation \eqref{eq:HInoice}  is the polarisation, $t_{\rm{tot}}$ is the total observing time, $N_d$ is number of dishes and $T_{\rm{sys}}$ is the sky temperature
\begin{eqnarray}
T_{\rm{sys}}(z) = 2.7 + 25 \left[\frac{400[ \rm{MHz}](1+z)}{\nu_{21}}\right]^{2.75} \,\, {\rm{K}} \,,
\end{eqnarray}
where $\nu_{21} = 1420 {\rm{MHz}}$  is the frequency of the 21 cm lime.  Table \ref{tab:survey} shows the parameters and the redshift range of these surveys.

%

\begin{table}[h]
	\centering
	\begin{tabular}{ | c | c | c | c | c | }
		\hline 
		Survey & 	Redshift range & $f_{\rm{sky}}$ & $t_{\rm{tot}} [\rm{hrs}]$ &  $N_{d}$\\
		\hline   \hline
		SKA1-MID Band 1  & 0.35-3.05  & 0.48 & 10000  &  197 \\ 
		\hline
	   H$\alpha$ emission line  & 0.7 - 2.0 & 0.35 & - & - \\
		\hline
	\end{tabular}
	\caption{Survey parameters for the HI intensity mapping survey~\cite{Ahmed:2019ocj} and H$\alpha$ emission line survey~\cite{Amendola:2016saw}.}
	\label{tab:survey}
\end{table}

\begin{figure}[h]
\centering 
\includegraphics[width=75mm,height=70mm] {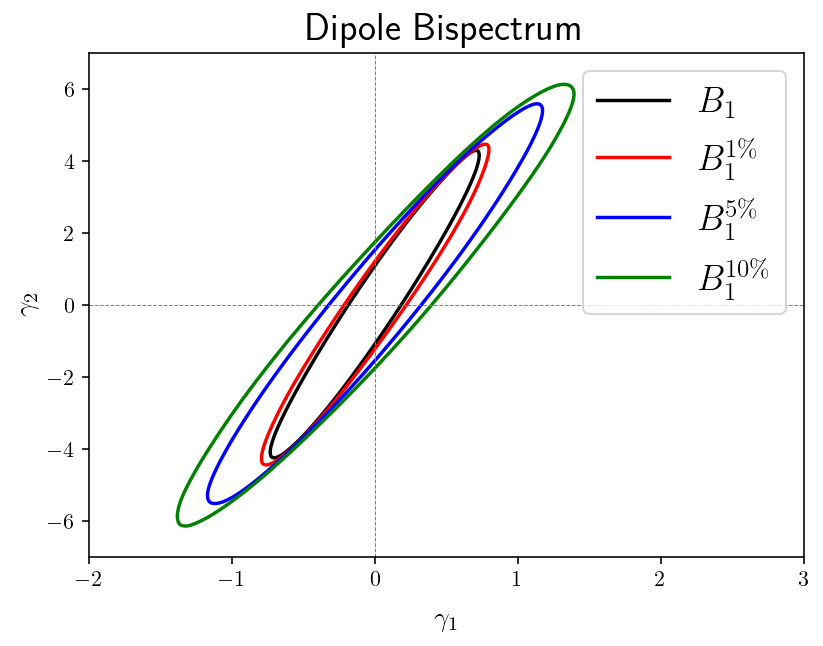}
\includegraphics[width=75mm,height=70mm] {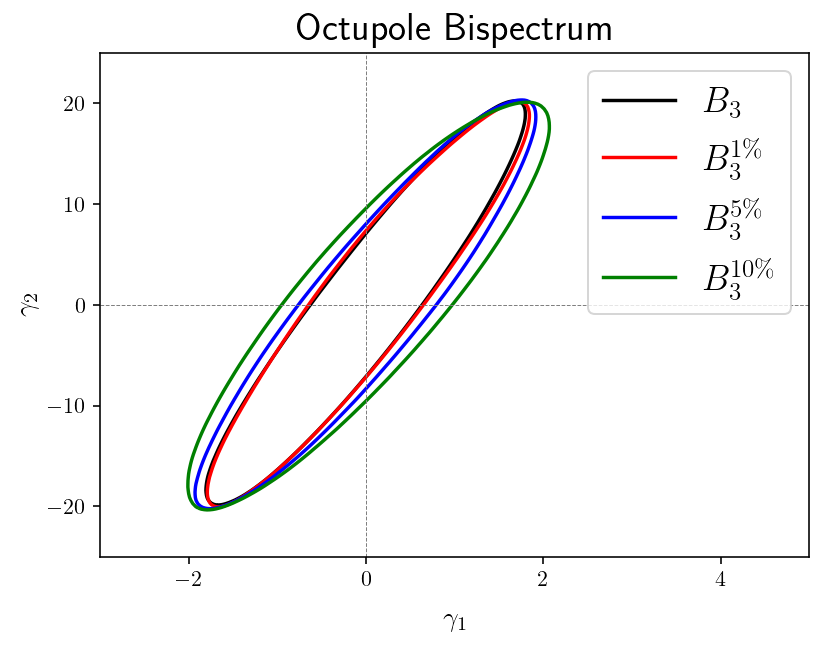}
\caption{\label{fig:Fisher1} {Left panel:  68\% confidence ellipse showing the constraint on  $\{\gamma_{1}, \gamma_{2}\}$ using the dipole of the H$\alpha$ galaxy  bispectrum with various priors on bias parameters. {Right panel: Just as it is in the left panel but for the octupole moment of the H$\alpha$ galaxy bispectrum.} }}
\end{figure}

We neglect the shot noise component of the total noise budget of the HI intensity mapping, as it is usually sub-dominant \cite{Santos:2015gra}. Also, we assume that foregrounds have been removed from the  HI intensity mapping signal. For recent developments of several foreground removal techniques, see~\cite{Cunnington:2019lvb}.
We show in Figure \ref{fig:SNR} the SNR for detecting the dipole moment of the galaxy cross-power spectrum for the H$\alpha$ emission line galaxy survey and  the HI intensity mapping survey, and the SNR for detecting the dipole and octupole of the H$\alpha$ galaxy bispectrum. 
Furthermore,  we forecast how well these surveys could  constrain  the parameters $\{\gamma_1, \gamma_{2}\}$  using the Fisher information matrix 
\begin{eqnarray}
F_{\alpha\beta}&=&\sum_{z_{\rm{min}}}^{z_{\rm{max}}}\sum_{T_{\rm{X}}}
\frac{\partial {X_{\ell}}(k_i)}{\partial \theta_\alpha}
{\rm Cov}^{-1}\left[X_{\ell}(k_i),X_{\ell}(k_j)\right]
\frac{\partial  {X}^{H}_{\ell}(k_j)}{\partial \theta_\alpha}.
\end{eqnarray}
It measures how steeply the likelihood falls as we move away from the best-fit model. The inverse of the Fisher information matrix approximates the  best possible covariance for measurement errors  on each parameter $\theta_{\alpha}$.  We calculate parameter constraint (marginal error) using  $ \sigma_{\alpha}^2 = \left( F^{-1}\right)_{\alpha\alpha}$.
The dipole moment of $P^{AB}_g$  is only sensitive to $\gamma_1$  at tree-level,  hence  we constrain $\theta_{\alpha} = \{\gamma_1\}$ only.  For  $P^{AB}_g$, we assume that the two surveys overlap in about 0.38 fraction of the sky \cite{Fonseca:2015laa}, hence our  Fisher matrix includes only the overlapping region~\cite{Viljoen:2020efi}
$F_{\alpha\beta} = F_{\alpha\beta} ^{AB} \left(\rm{overlap}\right) $. In principle, this is feasible since the aim of the  HI intensity mapping survey is to cover all sky by measuring the intensity of the redshifted 21cm line over the sky without the requirement to resolve individual galaxies~\cite{Santos:2015gra}. The left panel of figure 5 shows the constraint on $\gamma_1$ from $P_{AB}$ as a function of $z_{\rm{max}}$.

For the galaxy bispectrum we focus only on the H$\alpha$ emission line galaxy survey to constrain $\theta_{\alpha} = \{\gamma_1,\gamma_2\}$.  We do not use the bispectrum from HI intensity mapping because of concerns regarding the removal of the very dominant foregrounds; to date, HI intensity mapping has primarily been detected through cross-correlation with the galaxy count{~\cite{Masui:2012zc}. Attempts at finding reliable measurements of the auto-bispectrum of the HI intensity mapping is still in its infancy~\cite{Jolicoeur:2020eup}.}

{ Initially we assumed that the bias parameters would be determined precisely by the even multipole moments; however, they will not be determined with perfect accuracy.  Therefore, we consider how uncertainties in these parameters will affect the Fisher forecast analysis of the dipole and octupole moment of the $\Ha$ galaxy bispectrum.  To accomplish this, we parameterise the $\Ha$ bias parameters given in equations \eqref{eq:Halpha1} and \eqref{eq:Halpha2} as
\begin{eqnarray}
b^{\Ha}_{1}(z_i)&=& \mathcal{A}_{1} +  \mathcal{B}_{1} z_i\,,
\\
b^{\Ha}_{2}(z_i)&=& \mathcal{ A}_{2} + \mathcal{ B}_{2} z_i +  \mathcal{C}_{2} z^2_i +  \mathcal{D}_{2} z^3_i\,,
\end{eqnarray}
where we have introduced the following nuisance parameters
  \begin{eqnarray}
\theta_{\rm{nuisance}} &=&  \left\{ \mathcal{A}_1,  \mathcal{B}_1, \mathcal{A}_2,  \mathcal{B}_2, \mathcal{C}_2, \mathcal{D}_2\right\}\,.
\end{eqnarray}
We assume a local evolution for the tidal field, hence tidal bias parameter becomes $b^{\Ha}_{s^2}(z_i) = -{4} \left(b^{\Ha}_1(z_i) -1\right)/{7}$ \cite{Desjacques:2016bnm}.
We adopt the following fiducial values for the nuisance parameters $\theta_{\rm{nuisance}} $:
\begin{eqnarray}
\theta^{\rm{fid}}_{\rm{nuisance}} =  \left\{ 0.9,0.4,-0.704172,-0.207992,0.183023,-0.0007712
\right\}\,.
\end{eqnarray}
These are values are predicted by the halo model given in equations \eqref{eq:Halpha1} and \eqref{eq:Halpha2}.
}

\begin{figure}[h]
\centering 
\includegraphics[width=75mm,height=65mm] {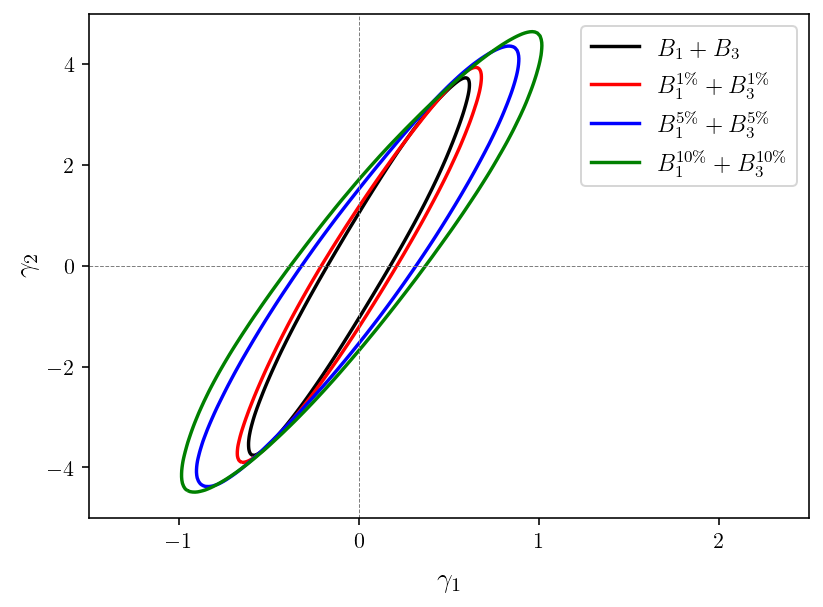}
\includegraphics[width=75mm,height=65mm] {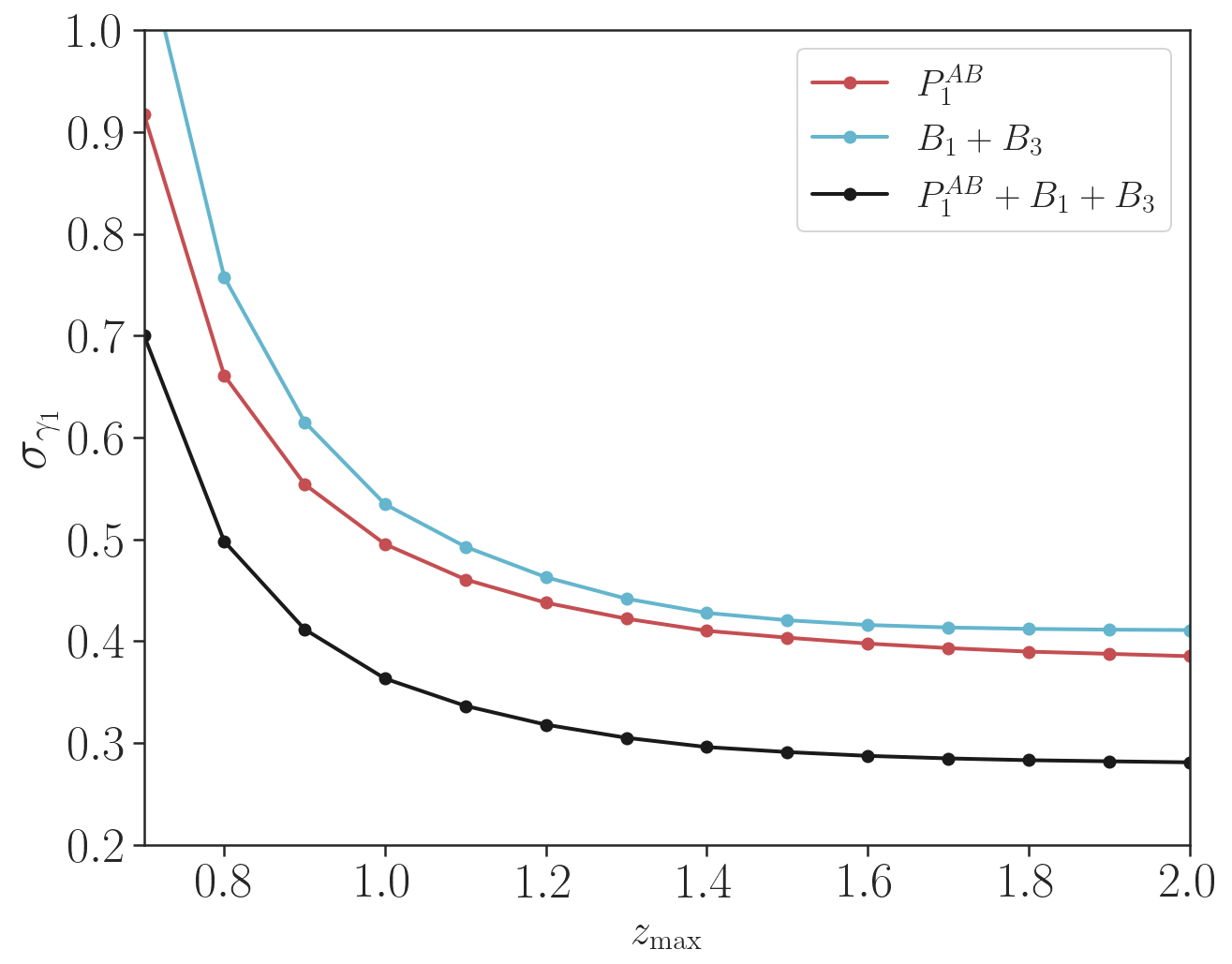}
\caption{\label{fig:combine} {{Left panel:}   68\% confidence ellipse  showing the constraint on  $\{\gamma_{1}, \gamma_{2}\}$ using both  dipole  and octupole multipole moments of the H$\alpha$ galaxy  bispectrum.
{Right panel:} The joint constraint on $\gamma_1$ from the combination of the dipole moment of the  galaxy cross-power spectrum and dipole + octupole moments of the H$\alpha$ galaxy bispectrum, after marginalising over $\gamma_2$. We also show the constraint on $\gamma_1$ from  combined dipole  and octupole moments  of the bispectrum, after marginalising over $\gamma_2$ and the nuisance parameters.}}
\end{figure}

We show in the right panel of figure \ref{fig:Fisher1}  the  $1\sigma$ confidence ellipse for $\{\gamma_1, \gamma_2\}$.  {We considered different Gaussian priors on the nuisance parameters and then marginalised over these; the results are shown in figure \ref{fig:Fisher1}. We recover the original constraint on $\{\gamma_1, \gamma_2\}$ as long as the nuisance parameters are determined to better than one percent accuracy.   This seems achievable from measurements of the even moments \cite{Tutusaus:2020xmc}. }
Finally, we  show in figure \ref{fig:combine} a joint constraint on $\gamma_1$ from the  combination of the dipole moment of $P^{AB}_{g1}$ and $B_{g1}$ . where we have marginalised over $\gamma_2$. Note that with the bispectrum of a single tracer alone we can constrain  {$\gamma_1 < { 0.28}$}. 

\begin{figure}[h]
\centering 
\includegraphics[width=75mm,height=75mm] {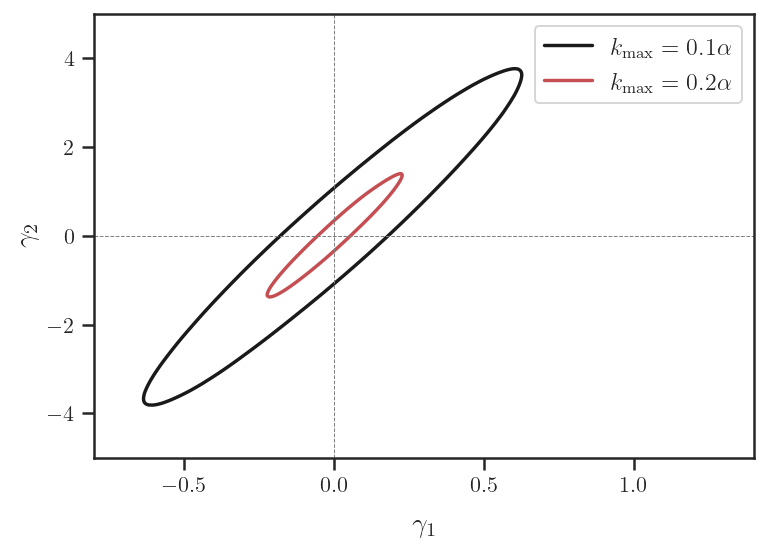}
\includegraphics[width=75mm,height=75mm] {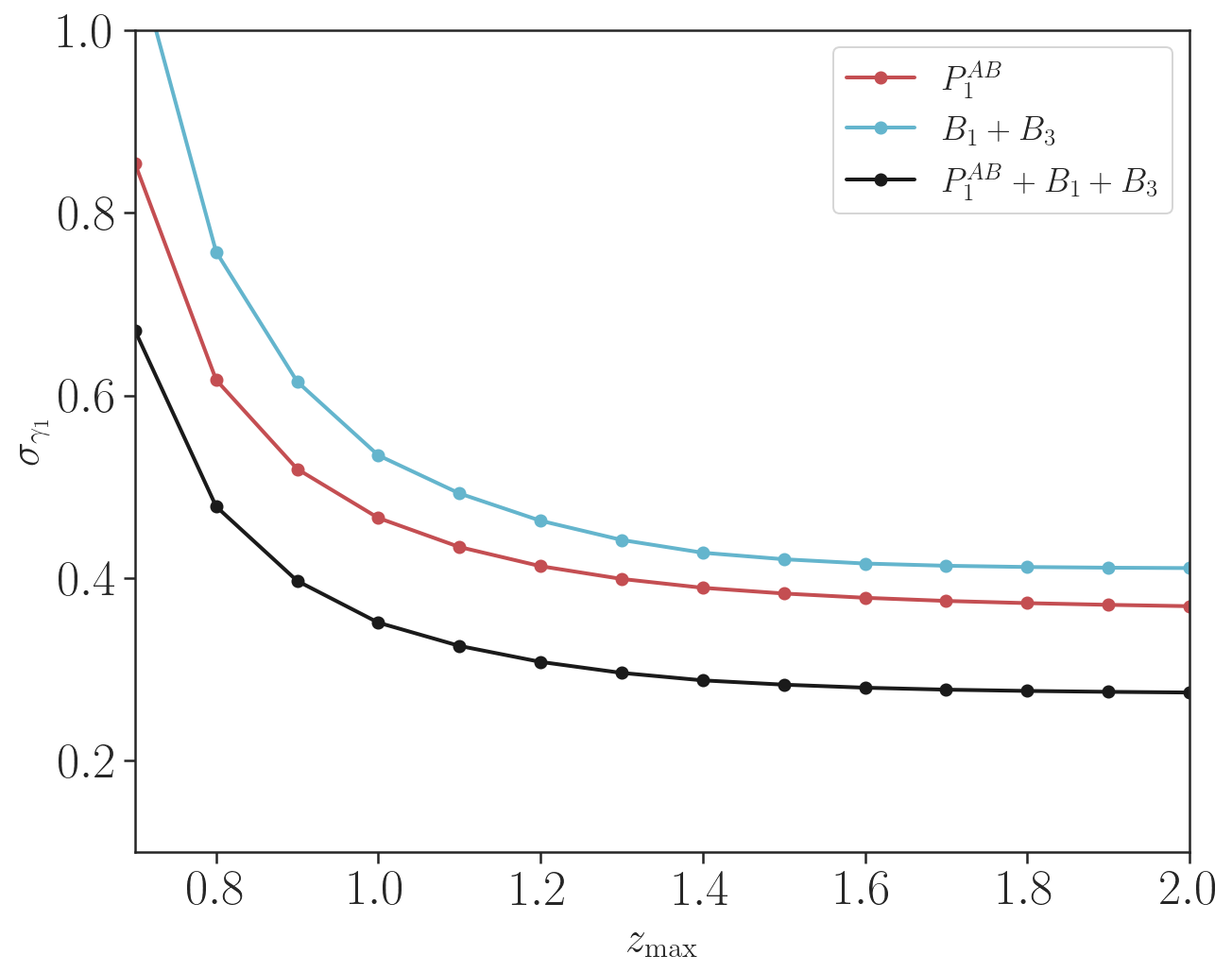}
\caption{\label{fig:Fisherappendix}{{Left panel: 68\% confidence ellipse  showing the dependence of the  constraint on  $\{\gamma_{1}, \gamma_{2}\}$ using  a combination of  dipole and  octupole for the H$\alpha$ galaxy  bispectrum on $k_{\rm{max}}$.  Right panel: The joint constraint on $\gamma_1$ from the combination of the dipole moment of the  galaxy cross-power spectrum and dipole + octupole moments of the H$\alpha$ galaxy bispectrum with $k_{\rm{max}} = 0.2\alpha$. We marginalised over $\gamma_2$ and fixed nuisance parameters to their predicted values. 
Here, $\alpha = (1+z)^{(2/(2+ n_s))} [h/{\rm{Mpc}}]$.} }}
\end{figure}

{The constraint on the galaxy-baryon relative velocity we reported depends sensitively on the choice of $k_{\rm{max}}$.  We have chosen a very conservative $k_{\rm{max}}$ motivated by the range of validity of the cosmological perturbation theory~\cite{Smith:2007sb}.   If, in the future, we are able to improve on the range of  modelling accuracy of the fluctuation of the number count of sources to  say  $k_{\rm{max}} = 0.2(1+z)^{(2/(2+ n_s))} [h{\rm{Mpc}^{-1}}]$, the  constraint on the galaxy-baryon relative velocity could improve significantly. See figure \ref{fig:Fisherappendix}. This motivates for further improvement in the modelling  of the Doppler contribution to the galaxy cross-power spectrum and bispectrum at higher k including 1-loop corrections~\cite{DiDio:2020jvo}. }

\section{Conclusion}\label{sec:conc}

We have explored in detail how the odd multipoles of the galaxy cross-power spectrum and bispectrum of the number count fluctuations could be used to test the equivalence principle on cosmological scales.  We developed this test by relaxing the assumption that the galaxy motion is geodesic (equivalence principle) on all scales in the derivation of the number count fluctuations beyond the Newtonian (Kaiser)  approximation.  Although the equivalence principle is one of the key principles of general relativity, there is no evidence that it applies to dark matter. 

The test we propose assumes that only the types of matter that have been confirmed to obey the equivalence principle at least on solar system scales (e.g. baryons) are geodesic~\cite{Touboul:2017grn}. {Our parametrisation assumes the velocity difference between galaxy and baryon is a scale-independent function of the galaxy velocity (equation  \eqref{eq:veldiffparam}); however, no further assumption was made about the motion of dark matter and the form of interaction in the dark sector.} This allowed us to express the violation of the equivalence principle in terms of the galaxy-baryon relative velocity. We parametrise the galaxy-baryon relative velocity in terms of the galaxy velocity, which is measurable.  {Furthermore, we assume that the  joint analysis of the even multipole moments of the galaxy power spectrum and bispectrum will be able to constrain the galaxy velocity, biases and other cosmological parameters to a much higher accuracy, for example see \cite{Tutusaus:2020xmc}\footnote{{The authors show that a cross-correlation analysis of the weak gravitational lensing information and the H$\alpha$ emission line galaxy clustering information from a Stage IV survey will be able to constrain the linear bias parameter to better than one percent accuracy.}}}, leaving the odd multipoles to constrain the equivalence principle.  

There are many mechanisms that can generate the relative velocity between baryons and cold dark matter in the Universe.  We enumerate a few: 
(1) Even for purely adiabatic initial perturbations, $v_{b\dm}$ is induced during baryon-photon decoupling;  the tight coupling of baryons to photon forces them to move  in a trajectory different  from that of the cold dark matter~\cite{Pitrou:2010ai}. In this case, it is usually assumed that given sufficient time after decoupling, the baryons will move to trace cold-dark matter but could leave a non-zero $v_{b\dm}$ at non-linear order~\cite{Barreira:2019qdl}.
(2) Isocurvature perturbations, potentially generated during inflation in a multi-field scenario, can set an initial condition for $v_{b\dm}$~\cite{Malik:2004tf,Carrilho:2018mqy}.
(3)  An interaction in the dark sector, such as where cold dark matter is coupled conformally and/or disformally to dark energy; such an interaction will boost $v_{b\dm}$ at late times.  We explore this possibility in greater detail in appendix \ref{sec:sourcesvbc}. 

We have shown that the Stage IV survey could constrain the galaxy-baryon relative velocity to {less than {28\%} of the galaxy velocity} using the cross-power spectrum and the bispectrum independently.
 Our analysis has aimed to be agnostic as to the origin of the galaxy-baryon relative velocity; however our choice of parameterisaton, given by equation \eqref{eq:Eqparametrization}, has implicitly assumed a late time violation of the equivalence principle. 
To compare to a theory, one would also need to relate the observed galaxy velocities to the velocities of their component parts; 
in the absence of velocity bias, the galaxy velocity is equal to the matter peculiar velocity $v_m$, the mass-weighted average of the cold dark matter and baryon velocities~\cite{Gleyzes:2015rua}:
$v_{g} =v_m =   x_{\dm} v_{\dm} + x_{b} v_{b},$
where $x_{\dm} = M_{\dm}/M_g$ and $x_b= M_b/M_g$.  Here, $M_{\dm}$ and $M_b$ are the masses of dark matter and baryon in the galaxy, respectively, and their sum $M_g = M_b + M_{\dm}$ is the total mass of the galaxy.  Therefore, $v_{gb} $ becomes  $v_{gb}  = x_{\dm} v_{\dm b} = - x_{\dm} v_{b\dm}$.

\section*{Acknowledgement}

We would like to thank Florian Beutler and Enea Di Dio  for discussions and clarifications on the covariance of the cross-power spectrum. 
Most of the tensor algebraic computations in this paper  were done with the tensor algebra software xPand \cite{Pitrou:2013hga} which is based on xPert~\cite{Brizuela:2008ra}.
OU, KK and RC  are supported by the UK STFC grant ST/S000550/1.  KK is also supported by the European Research Council under the European Union's Horizon 2020 programme (grant agreement No.646702 ``CosTesGrav"). 


\appendix

\section{The source of baryon-cold dark matter relative velocity}\label{sec:sourcesvbc}

\subsection{Interacting dark sector in scalar tensor theory}

We discuss how a possible interaction in the dark sector could provide a source for the baryon-dark matter relative velocity and investigate the number of parameters that are required to characterise its evolution. We study a general action for a Scalar-Tensor theory, where the gravitational action is a sum of the action for the quintessence scalar field, $S_{\phi}$, minimally coupled to the Einstein-Hilbert action, $S_{\rm{EH}}$,  in the presence of standard matter, $S_{\rm{M}}$, and the dark matter field, $S_{\dm} $,
 \begin{eqnarray}\nonumber 
S &=& \sum_{I} S_{I} \\  \label{eq:full-action}
&=& S_{\rm{EH}}[g_{\mu\nu}] + S_{\rm{M}}[{\rm{std~matter}},g_{\mu\nu}] +S_{\phi}[\phi,g_{\mu\nu}]+ S_{\dm}[{\rm{dark~matter}},\tilde{g}_{\mu\nu}]\,,
\end{eqnarray}
where   $S_{\dm}$  and $S_g=S_{\rm{EH}}+ S_{\rm{M}}+S_{\phi}$ are given by~\cite{vandeBruck:2017idm}
\begin{eqnarray}
S_{\dm} [\tilde{g}]&=&\int \mathrm{d}^4 x \sqrt{-\tilde{g}} \, \mathcal{L}_{\dm}\left(\tilde{g}_{\mu\nu},\varphi\right)\,,
\\
S_{\rm{g}}[g] &=& \int \mathrm{d}^4 x \sqrt{-g}\left[ \frac{1}{2\kappa^2} R - \frac{1}{2}g^{\mu\nu}\partial_{\mu}\phi \partial_{\nu}\phi - V(\phi) + \mathcal{L}_\mathrm{M} \right]  \, .
\end{eqnarray}
Here $\varphi$ is a dark matter field, which sees the metric $\tilde{g}_{\mu\nu}$, and $\mathcal{L}_{\dm}$ is its Lagrangian while $\mathcal{L}_\mathrm{M} $ is the Lagrangian for the standard matter fields, which instead see the metric ${g}_{\mu\nu}$. The scalar field $\phi$, has a canonical kinetic term with a potential $V(\phi)$ and $\kappa^{-1} \equiv [8\pi G]$ is related to the Planck mass.

We consider a scenario where  $\tilde{g}_{\mu\nu}$ is related to $g_{\mu \nu}$ as 
\begin{eqnarray}\label{eq:gravitationalcoupling}
\tilde{g}_{\mu\nu} = C(\phi) g_{\mu\nu} + D(\phi) \partial_{\mu}\phi \partial_{\nu}\phi\,,
\end{eqnarray}
where $C(\phi)$ and $D(\phi)$ are conformal and disformal coupling functions respectively. These functions can also depend on the kinetic term $X = -\partial_{\mu}\phi \partial_{\nu}\phi/2$, however, we neglect this dependence going forward for simplicity. In the fluid limit, we can define the energy momentum tensors, $T^{\mu\nu}_{\! I}$ , associated with each of the fields in equation \eqref{eq:full-action} and parameterise each one in terms of the energy density and pressure in its frame of reference according to 
\begin{eqnarray}\label{eq:EMT}
T^{\mu\nu}_{\! I} = \left(\rho_{\!_I}+P_{\!_I}\right)u^\mu_{\! I} u^\nu_{\! I} + P_{\!_I}g^{\mu\nu}\,.
\end{eqnarray}
Here $I$ indicates the type of matter; $I = {\rm{c}}$ for the  dark matter, $I = \phi$ for the quintessence scalar  field and $I = {\rm{M}}$ standard matter. 

Each field obeys the following energy-momentum conservation equations
\begin{eqnarray}
\nabla^{\mu}T_{\mu\nu}^{\rm{M}} &=&0 \,,
\label{eq:conservation_equation-for-sm}
\\
\nabla^{\mu} T^{\dm}_{\mu\nu} &=& Q\left(\phi,u^{\mu}\nabla_{\mu}\phi,\rho_{\rm{c}}\right) \nabla_{\nu}\phi\,.
\label{eq:conservation_equation-for-dm}
\end{eqnarray}
The function $Q$ is given by~\cite{Mifsud:2017fsy}
\begin{eqnarray}
Q\left(\phi,u^{\mu}\nabla_{\mu}\phi,\rho_{\dm}\right)&=& \frac{1}{2}\left\{ \frac{\d \ln C}{\d \phi} T_{\dm} + \frac{D}{C} \frac{\d \ln D}{\d \phi} T^{\mu\nu}_{\dm} \nabla_{\mu}\phi \nabla_{\nu}\phi - 2\nabla_{\mu}\left[ \frac{D}{C} T^{\mu\nu}_{\dm} \nabla_{\nu}\phi\right]\right\}\,,
\end{eqnarray}
where $T_{\rm{\dm}}$ is the trace of the dark matter energy-momentum tensor.

\subsection{Interacting dark sector in cosmological perturbation theory}
At linear order in cosmological perturbation theory and in the weak field limit, the continuity equations for baryons and dark matter from equations \eqref{eq:conservation_equation-for-sm} and \eqref{eq:conservation_equation-for-dm} are 
\begin{eqnarray}
{\delta\one_{b}}'  &=&- \partial_i\partial^i v_{b}\one \,,
\label{eq:continuitybaryoneqn1}
\\
{\delta\one_{\dm}}'  &=&- \partial_i\partial^i v_{\dm}\one -\HH\Theta_1 \left(1- \Theta_3 \right)\delta_{\dm}\one\,,
\label{eq:continuityeqn1}
\end{eqnarray}
and the Euler equations are
\begin{eqnarray}
{\partial^i v_{b}\one}' &+& \HH \partial^iv_{b}\one+ {\partial^i}\Phi\one=0\, , \label{eq:EulereqnBaryons1}
\\
{\partial^i v_{\dm}\one}' &+& \HH \left[1+\Theta_1\right] \partial^iv_{\dm}\one+\left[1+\Theta_2\right] {\partial^i}\Phi\one=0\, ,
\label{eq:ModefiedEulereqn1}
\end{eqnarray}
In \cite{Bonvin:2018ckp}, the authors assumed that $v_{g}\one = v_{\dm}\one$,  then made use of equation \eqref{eq:ModefiedEulereqn1} to relate the gravitational potential to $v_{\dm}\one$. 
Here we have introduced the parameterizations following \cite{Bonvin:2018ckp}
\begin{eqnarray}
\Theta_1&=&-\frac{\bar{\phi}'}{\HH}\frac{\bar{Q}}{\bar{\rho}_{\dm}}\,,
\qquad
\Theta_2=-\frac{\bar{Q}^2}{\kappa\bar{\rho}_{\dm}}
\qquad \Theta_3 = \frac{\partial \ln \bar{Q}}{\partial \ln \rho_{\dm}}\,.
\end{eqnarray}
These parameters parametrise the violation of the equivalence principle due to the fact that dark matter moves in a geodesic that is different from that of the standard matter because of the interaction with the quintessence scalar field through the conformal or the disformal coupling. This shows that we need three free parameters to describe the evolution of the dark matter density and velocity at linear order. 

Similarly at the second order, the continuity equations for baryons and dark matter are given by 
\begin{eqnarray}
{\delta\two_{b}}'   &=&- \partial_i\partial^i v_{b}\two
+2\delta\one_{b}  \partial_i\partial^i v_{b}\one  + 2\partial_i v_{b}\one\partial^i\delta_{b}\one 
+ \mathcal{O}\left((\partial\Phi\one)^2\right)\,,\label{eq:contuniutyeqnsm}
\\
{\delta\two_{\dm}}'   &=&- \partial_i\partial^i v_{\dm}\two
-\HH\bigg[\Theta_1 \left(1-\Theta_3 \right)\delta_{\dm}\two\bigg]+2\delta\one_c  \partial_i\partial^i v_{\dm}\one  + 2\partial_i v_{\dm}\one\partial^i\delta_{\dm}\one 
\\ \nonumber &&
+ \Theta_1 \Theta_4(\delta_{\dm}\one)^2 + \mathcal{O}\left((\partial\Phi\one)^2\right)\,.
\label{eq:contuniutyeqndm}
\end{eqnarray}
The corresponding Euler equation become
\begin{eqnarray}
\partial_i {v_{b}\two}' & +& \HH\partial_i {v_{b}\two}+   \partial_i\Phi\two  + 2 \partial_i \partial_j v_{b}\one \partial^jv_{b}\one   
 +\mathcal{O}(\Phi\one\partial\Phi)= 0\,, \label{eq:EulereqnBaryons2}
\\ 
\partial_i {v_{\dm}\two}' &+& \HH\left[1 + \Theta_1\right]\partial_i {v_{\dm}\two}+  \left[1 + \Theta_2\right]\partial_i\Phi\two  + 2 \partial_i \partial_j v\one \partial^jv\one   
\\ \nonumber &&
 + 2\HH\Theta_1 \left( \Theta_3 -1\right)\delta_{\dm}\one\partial_i v_{\dm}\one  -2\Theta_2\left( 1  -\Theta_3\right)\delta_{\dm}\one{\partial_i  \Phi\one}
 +\mathcal{O}(\Phi\one\partial\Phi)
= 0 \, ,
\label{eq:Eulereqndm2}
\end{eqnarray}
where we have introduced yet another parameter to describe the self-coupling strength of the dark matter density field
\begin{eqnarray}
\Theta_4 = \frac{1}{Q}\frac{\partial^2 Q}{(\partial \ln \rho_{\dm})^2} \,.
\end{eqnarray}
The second order dark matter density and velocity can be obtained by solving these equations. Note that, since the gravitational potential is sourced by the dark matter and baryon density, the equations for baryons and cold dark matter are coupled.   

We have shown that we need four free parameters $ \big\{ \Theta_{1} , \Theta_{2} , \Theta_{3} , \Theta_{4} \big\}$ in order to characterise the interaction between dark matter and baryons in this model. By specifying these parameters, we can compute the relative velocity between baryons and dark matter, and predict the equivalence principle violation parameters $\Upsilon_{1,2}$ and $\beta_{1,2}$.


\providecommand{\href}[2]{#2}\begingroup\raggedright\endgroup

\end{document}